\documentclass[a4paper,11pt]{article}
\pdfoutput=1 

\usepackage{jcappub} 

\usepackage{graphicx}
\usepackage{dcolumn}
\usepackage{bm}
\usepackage{hyperref}
\usepackage{color}

\usepackage{amsmath}
\usepackage{acronym}
\usepackage{xspace}
\usepackage{subfigure}

\usepackage{tabularx}
\newcolumntype{L}[1]{>{\hsize=#1\hsize\raggedright\arraybackslash}X}%
\newcolumntype{R}[1]{>{\hsize=#1\hsize\raggedleft\arraybackslash}X}%
\newcolumntype{C}[1]{>{\hsize=#1\hsize\centering\arraybackslash}X}%

\definecolor{olive}{rgb}{0.5, 0.5, 0.0}

\newcommand{\kmsMpc}{\ensuremath{\mbox{km s}^{-1} \,\mbox{Mpc}^{-1}}\xspace}
\newcommand{\sqdeg}{\ensuremath{\text{deg}^2}\xspace}
\newcommand{\gwcosmo}{\textsc{gwcosmo}\xspace}
\newcommand{\icarogw}{\textsc{Icarogw}\xspace}

\newcommand{\dynesty}{\textsc{dynesty}\xspace}

\def\ie{{\emph{i.e.~}}}
\def\eg{{\emph{e.g.~}}}
\newcommand{\BBHpopempty}{$33^{+51}_{-18}$ \kmsMpc}
\newcommand{\combinedpoppluscatalogue}{$69^{+12}_{-7}$ \kmsMpc}
\newcommand{\oldgwcosmo}{$68^{+8}_{-6}$ \kmsMpc}

\acrodef{H0}[$H_0$]{the Hubble constant}
\acrodef{LVC}[LVC]{LIGO and Virgo Collaborations}
\acrodef{GW}[GW]{gravitational wave}
\acrodef{EM}[EM]{electromagnetic}
\acrodef{CMB}[CMB]{cosmic microwave background}
\acrodef{SN}[SN]{supernovae}
\acrodef{MDC}[MDA]{mock data analysis}
\acrodefplural{MDC}[MDAs]{mock data analyses}
\acrodef{BNS}[BNS]{binary neutron star}
\acrodef{NS}[NS]{neutron star}
\acrodef{BBH}[BBH]{binary black hole}
\acrodef{BH}[BH]{black hole}
\acrodef{LCDM}[$\Lambda$CDM]{$\Lambda$-cold-dark-matter}
\acrodef{SNR}[SNR]{signal-to-noise ratio}
\acrodef{F2Y}[F2Y]{First Two Years}
\acrodef{GRB}[GRB]{gamma-ray burst}
\acrodef{SH0ES}[SH0ES]{Supernovae, $H_0$, for the Equation of State of Dark energy}
\acrodef{O1}[O1]{first observing run}
\acrodef{O2}[O2]{second observing run}
\acrodef{O3}[O3]{third observing run}
\acrodef{NSBH}[NSBH]{neutron star - black hole pair}
\acrodef{CBC}[CBC]{compact binary coalescence}
\acrodef{O2H0}[O2-$H_0$]{O2 Hubble constant}
\acrodef{PSD}[PSD]{power-spectral-density}
\acrodefplural{PSD}[PSDs]{power-spectral-densities}
\acrodef{ra}[RA]{right ascension}
\acrodef{dec}[dec]{declination}
\acrodef{dl}[$d_L$]{luminosity distance}
\acrodef{LOS}[LOS]{line-of-sight}
\acrodef{KDE}[KDE]{kernel density estimate}
\acrodef{FAR}[FAR]{false alarm rate}
\acrodef{IFAR}[IFAR]{inverse false alarm rate}
\acrodef{PE}[PE]{parameter estimation}
\acrodef{DES}[DES]{Dark Energy Survey}
\acrodef{SDSS}[SDSS]{Sloan Digital Sky Survey}
\acrodef{GLADE}[GLADE]{Galaxy List for the Advanced Detector Era}
\acrodef{GWENS}[GWENS]{Gravitational Wave Events in Sloan}
\acrodef{LSC}[LSC]{LIGO Scientific Collaboration}
\acrodef{mth}[$m_\text{th}$]{apparent magnitude threshold}
\acrodef{GWTC-3}[GWTC-3]{the third Gravitational-Wave Transient Catalogue}
\acrodef{SFR}[SFR]{star formation rate}

\title{Joint cosmological and gravitational-wave population inference using dark sirens and galaxy catalogues}

\author[a,b,1]{Rachel Gray,}
 \emailAdd{rachel.gray@glasgow.ac.uk}
\affiliation[a]{SUPA, University of Glasgow, Glasgow, G12 8QQ, United Kingdom}
\affiliation[b]{Department of Physics and Astronomy, Queen Mary University of London, Mile End Road, London, E1 4NS, United Kingdom}

\author[c]{Freija Beirnaert,}
\affiliation[c]{Ghent University, Proeftuinstraat 86, Ghent, Belgium}

\author[d]{Christos Karathanasis,}
\affiliation[d]{Institut de Física d’Altes Energies (IFAE), Barcelona Institute of Science and Technology, Barcelona, Spain}

\author[e,f]{Beno\^it Revenu,}
\affiliation[e]{Subatech, CNRS, Institut Mines-Telecom Atlantique, Nantes Universit\'e, France}
\affiliation[f]{Universit\'e Paris Cit\'e, CNRS, Astroparticule et Cosmologie, F-75013 Paris, France}

\author[c]{Cezary Turski,}

\author[b]{Anson Chen,}

\author[b,g]{Tessa Baker,}
\affiliation[g]{Institute of Cosmology and Gravitation, University of Portsmouth, Burnaby Road, Portsmouth PO1 3FX, UK}

\author[h]{Sergio Vallejo,}
\affiliation[h]{Instituto de F\'isica, Universidad de Antioquia, A.A.1226, Medell\'in, Colombia}

\author[h]{Antonio Enea Romano,}

\author[i]{Tathagata Ghosh,}
\affiliation[i]{Inter-University Centre for Astronomy and Astrophysics, Post Bag 4, Ganeshkhind, Pune 411 007, India}

\author[c]{Archisman Ghosh,}

\author[j]{Konstantin Leyde,}
\affiliation[j]{Universit\'e Paris Cit\'e, CNRS, Astroparticule et Cosmologie, F-75013 Paris, France}

\author[k]{Simone Mastrogiovanni,}
\affiliation[k]{INFN, Sezione di Roma, I-00185 Roma, Italy}

\author[i]{Surhud More}

\date{\today}

\abstract{In the absence of numerous gravitational-wave detections with confirmed electromagnetic counterparts, the ``dark siren'' method has emerged as a leading technique of gravitational-wave cosmology. The method allows redshift information of such events to be inferred statistically from a catalogue of potential host galaxies.
Due to selection effects, dark siren analyses necessarily depend on the mass distribution of compact objects and the evolution of their merger rate with redshift. Informative priors on these quantities will impact the inferred posterior constraints on the Hubble constant ($H_0$). It is thus crucial to vary these unknown distributions during an $H_0$ inference. This was not possible in earlier analyses due to the high computational cost, restricting them to either excluding galaxy catalogue information, or fixing the gravitational-wave population mass distribution and risking introducing bias to the $H_0$ measurement.
This paper introduces a significantly enhanced version of the Python package \gwcosmo, which allows joint estimation of cosmological and compact binary population parameters. 
This thereby ensures the analysis is now robust to a major source of potential bias. The gravitational-wave events from the Third Gravitational-Wave Transient Catalogue are reanalysed with the GLADE+ galaxy catalogue, and an updated, more reliable measurement of $H_0=$ \combinedpoppluscatalogue is found (maximum a posteriori probability and 68\% highest density interval).
This improved method will enable cosmological analyses with future gravitational-wave detections to make full use of the information available (both from galaxy catalogues and the compact binary population itself), leading to promising new independent bounds on the Hubble constant.}

\begin{document}

\maketitle
\flushbottom


\section{Introduction}

The disagreement surrounding the value of \ac{H0} is one of the dominant unanswered questions in current cosmology. 
Current measurements of \ac{H0} are in tension with each other to a level above $4\sigma$ (see, \eg \cite{Aghanim:2018eyx,Riess_2022}), with local (low-redshift) measurements typically preferring a higher value, and early-universe (high-redshift, model-dependent) measurements preferring a lower one. The cause of this tension is unknown, but potential explanations include 
unaccounted-for measurement errors (which are more likely to impact the local measurements), or a flaw in the current standard cosmological model, \ac{LCDM} (which would impact the early-universe measurements) \cite{2019NatAs...3..891V,DiValentino_2021}. 

The relationship between redshift $z$, luminosity distance $d_L$, and \ac{H0} can be expressed as
\begin{equation}
    d_L = \dfrac{c(1+z)}{H_0} \int^z_0 \dfrac{dz'}{\sqrt{(1+z')^3 \Omega_M + \Omega_\Lambda}},
\end{equation}
for a flat universe. At low redshifts this simplifies to
\begin{equation}
    d_L \approx \dfrac{cz}{H_0} .
\end{equation}
Hence, for a local measurement of \ac{H0} to be made, all that is required is redshift and luminosity distance information.  
The amplitude of a gravitational wave signal from a \ac{CBC} directly encodes the luminosity distance to the source, requiring no external calibration. This property has led to these types of \ac{GW} events being commonly termed ``standard sirens''. If information about the redshift of the source can be supplied, the \ac{GW} data can be used to produce a measurement of \ac{H0} which is independent of the cosmic distance ladder, and of all \ac{EM} measurements of \ac{H0} to date. Herein lies significant potential for the field of gravitational wave astronomy to resolve the long-standing Hubble tension.  

If a uniquely-identified \ac{EM} counterpart can be associated to the \ac{GW} event (making it a ``bright siren''), the redshift of the \ac{GW} host galaxy can be determined via spectroscopy, allowing for a direct measurement of \ac{H0}. Thus far this has only been possible for the detection of the binary neutron star GW170817 and its \ac{EM} counterpart \cite{GW170817:discovery,GW170817:MMA}. 
The small localisation region of the event (possible because the two LIGO detectors \cite{Aasi_2015} and Virgo detector \cite{Acernese_2015} were online at the time) meant that multiple telescopes were able to search the relevant area over a short time-frame, leading to the identification of the kilonova associated with the \ac{GW} event, and subsequently the identification of its host galaxy. This led to the first ever measurement of \ac{H0} from \acp{GW} \cite{GW170817:H0}, a remarkably informative measurement from a single event.

Since then, the third observing run of Advanced LIGO, Virgo and KAGRA \cite{10.1093/ptep/ptaa125} has taken place, with observations lasting for just under a year and increasing the current catalogue of \ac{GW} transients to $\sim90$ \cite{GWTC-3}. Despite improvements in the sensitivity of the LIGO and Virgo detectors \cite{PhysRevD.102.062003,PhysRevLett.123.231108}, no \ac{EM} counterparts have been uniquely associated with any of the subsequent events. This distinct lack of \ac{EM} counterparts has led to the blossoming of methods for supplying redshift information to these ``dark sirens''. 

Current dark sirens methods can be broadly split into two further categories:\footnote{Here we make a distinction between dark siren methods and the cross-correlation method, in which \acp{GW} are cross-correlated with galaxy surveys or other tracers in order to obtain cosmological constraints. For details on the cross-correlation method see, \eg \cite{PhysRevD.93.083511,2018arXiv180806615M,2020MNRAS.494.1956M,2021PhRvD.103d3520M,2022MNRAS.511.2782C,2020ApJ...902...79B,2022arXiv220303643M}.}
\begin{itemize}
    \item The galaxy catalogue method: here the redshift information is provided by a catalogue of potential host galaxies within the sky localisation region of a \ac{GW} event (see \cite{Schutz:1986,DelPozzo:2012,Fishbach:2018edt,Soares-Santos:2019irc,Gray:2019ksv,GW190814:DES,Finke:2021aom,Gray2022}). Each galaxy contributes to a hypothetical measurement of \ac{H0}, such that the galaxy structure within a \ac{GW} event's localisation volume is reflected in the \ac{H0} posterior it produces. How informative the individual events are will depend strongly on their localisation volumes. By combining the contributions of many events, the true value of \ac{H0} will be measured as other values will statistically average out.
    \item The population method, also known ``spectral sirens'': here features in the mass distribution of the compact object population break the mass-redshift degeneracy 
    (see \cite{1993ApJ...411L...5C,Taylor:2011fs,2019ApJ...883L..42F,PhysRevD.104.062009,PhysRevLett.129.061102}). This method can be thought of as analogous to the redshifting of emission lines in \ac{EM} spectra: if the location of a feature is known in the source frame, its measured position in the detector frame -- shifted by a factor of $(1+z)$ -- informs us about the redshift of the source. 
\end{itemize}
These two varieties of dark sirens methods (as presented in \cite{Gray2022,PhysRevD.104.062009}) were applied in the most recent LIGO-Virgo-KAGRA cosmology paper \cite{GWTC-3:cosmology}, making use of the \ac{GW} data from \ac{GWTC-3} \cite{GWTC-3}. 
Until recently, the two dark sirens methods have always been carried out separately due to the computational difficulty in incorporating galaxy catalogue information (which tends to be highly discrete) into a sampling method, and the computational cost of incorporating a reliable correction for galaxy catalogue incompleteness. This means that the population method has not benefited from the additional constraining power of galaxy redshifts. Conversely, the galaxy catalogue method required fixing the \ac{GW} population when computing the selection function, which opens it up to potential bias if the assumed model (with fixed hyper-parameters) is incorrect. This second point is clearly more serious, as it could lead to a significant bias in the measurement of \ac{H0} if not corrected.

This paper presents a novel method for carrying out cosmological and \ac{GW} population parameter estimation jointly, such that the resulting posterior on \ac{H0} is free from bias due to \ac{GW} population assumptions.\footnote{Of course this assumes that the correct population models have been assumed. See, \eg \cite{PhysRevD.104.062009}.} This method combines the best of both dark siren techniques above: for the large numbers of \acp{BBH} at high redshifts the galaxy catalogues are relatively uninformative (due to incompleteness), but these events offer significant constraining power through their mass distribution. Near-by, well-localised \acp{BNS}, \acp{NSBH} and \acp{BBH} are relatively lower in numbers and hence do not strongly probe the mass distribution, but instead can access the highly informative nature of a galaxy catalogue to provide redshift information. This updated method is implemented within the Python package \gwcosmo,\footnote{\url{https://git.ligo.org/lscsoft/gwcosmo}} and used to reanalyse the \ac{GWTC-3} data in combination with the GLADE+ galaxy catalogue \cite{GLADE+}, leading to a final updated value of \ac{H0} which utilises 47 \ac{GW} events and incorporates redshift information from the \ac{BH} mass distribution, a full-sky galaxy catalogue, and the one confirmed \ac{EM} counterpart to date. 

This version of \gwcosmo, which is about 1000 times faster than its predecessor, is well suited to the challenges that lie ahead in the field of \ac{GW} cosmology, in terms of the large numbers of gravitational wave events which are expected over the coming years, and the increasing depths of galaxy surveys which means that the sheer quantity of data that will be available will require efficient handling and analyses which can scale to large datasets. Looking to the future, beyond the \ac{H0} tension, this version of \gwcosmo can additionally be used to put constraints on models of modified gravitational-wave propagation \cite{AChen2023}.

Section \ref{sec:method} presents the updated methodology in a Bayesian framework, including introducing the crucial concept of a ``\ac{LOS} redshift prior''. Section \ref{sec:data} discusses the choice of \ac{GW} data and galaxy catalogues, including how the GLADE+ galaxy catalogue is converted into a \ac{LOS} redshift prior. Section \ref{sec:results} reanalyses the GWTC-3 data with the updated methodology, under the same assumptions as \cite{GWTC-3:cosmology}, and quantifies the impact of various assumptions, then presents a full population + galaxy catalogue reanalysis of GWTC-3, including an updated measurement of \ac{H0} which utilises the greatest amount of \ac{GW} information to date. Finally, section \ref{sec:conclusions} concludes the paper and discusses the application of this method to future data.

Finally, a paper presenting an alternative method of joint cosmological and population inference using the \icarogw package was released while this paper was in preparation \cite{Mastrogiovanni2023}. The methodology is derived using a different approach (expressing the likelihood in terms of \acp{CBC} merger rates) which is consistent with the \gwcosmo methodology where equivalent assumptions are made. The way in which gravitational-wave data is handled between the two codes differs, as \icarogw computes numerical integrals involved in the evaluation of the likelihood by summing over posterior samples, while \gwcosmo constructs a \ac{KDE} for different lines of sight. The \icarogw approach enables a faster analysis, at the expense of requiring that pixel sizes and galaxy redshift distributions are large enough to contain a representative number of posterior samples. The \gwcosmo approach, while potentially slower, does not suffer from any of these resolution effects due to the smoothing of the \ac{GW} posterior.\footnote{No formal timing tests have yet been carried out between the two codes, so a definitive comparison is left for future work.} Differences aside, we find good agreement with their results when considering the same set of 42 \acp{BBH} for a full catalogue+population analysis, and when considering the highly informative dark siren GW190814 for a catalogue-only analysis.


\section{Method\label{sec:method}}

In general, the posterior on a set of hyper-parameters $\Lambda$, given a set of \ac{GW} data $\{x_\text{GW}\}$, corresponding to $N_\text{det}$ detections, $\{D_\text{GW}\}$, is given by \cite{Mandel:2018mve,Vitale:2020aaz}:
\begin{equation}
\begin{aligned}
p(\Lambda|\{x_\text{GW}\},\{D_\text{GW}\}) &\propto p(\Lambda) p(N_\text{det}|\Lambda)  \prod^{N_\text{det}}_i p(x_{\text{GW}i}|D_{\text{GW}i},\Lambda).
\end{aligned}
\end{equation}
The parameter $D_{\text{GW}i}$ is a binary parameter (taking the value 0 or 1) denoting that the \ac{GW} data of event $i$ passed some detection threshold statistic in order to be deemed ``detected''. The term $\Lambda$ includes the cosmological parameters of interest (nominally \ac{H0} for this analysis) as well as a set of population hyper-parameters which describes the underlying mass distribution of compact binaries and their distribution in redshift.
This expression expands to:
\begin{equation}\label{Eq:sum_likelihood_compact}
\begin{aligned}
p(\Lambda|\{x_\text{GW}\},\{D_\text{GW}\},I) &\propto p(\Lambda|I) p(N_\text{det}|\Lambda,I) \prod^{N_\text{det}}_i \dfrac{\int p(x_{\text{GW}i}|\theta,\Lambda,I)p(\theta|\Lambda,I) d\theta}{\int p(D_{\text{GW}i}|\theta,\Lambda,I)p(\theta|\Lambda,I) d\theta},\\
&\propto p(\Lambda|I) p(N_\text{det}|\Lambda,I) \left[\int p(D_\text{GW}|\theta,\Lambda,I)p(\theta|\Lambda,I) d\theta \right]^{-N_\text{det}}\\ &\hspace{100pt}\times \prod^{N_\text{det}}_i \int p(x_{\text{GW}i}|\theta,\Lambda,I)p(\theta|\Lambda,I) d\theta,
\end{aligned}
\end{equation}
where it's been written explicitly that the individual \ac{GW} likelihood need to be marginalised over individual event parameters (such as sky location, masses, inclination, spins, redshift etc.), grouped together and denoted by $\theta$. \ac{GW} selection effects are incorporated in the term $p(D_\text{GW}|\Lambda) = \int p(D_\text{GW}|\theta,\Lambda)p(\theta|\Lambda) d\theta$. This is the ``probability of detection'' for a \ac{CBC} drawn from a population described by parameters $\Lambda$, and because it applies to the whole population of \acp{CBC} (as opposed to being event-specific) it can be taken outside the product.\footnote{The term $p(D_{\text{GW}i}|x_{\text{GW}i},\Lambda,I)$, which appears when you first expand the expression using Bayes theorem, is equal to 1 as the GW data is, by definition, detected.} The term $I$ on the right-hand side of each expression is a catch-all term which holds any additional assumptions that have not been explicitly expressed.

The method, as written to this point, is the same between the population and the galaxy catalogue methods of computing \ac{H0}, but at this point it is worth reiterating the manner in which these analyses will differ: the prior redshift information. In the population method, the redshift prior would commonly be taken to be uniform in comoving volume (in order to model the galaxy distribution with the least informative assumptions when lacking any galaxy positional data), modified by some additional term for the evolution of the merger rate of binaries with redshift. In this scenario, the redshift information necessary for cosmological inference comes mainly from the distribution of the observed \ac{GW} masses $p(m^d_{1}, m^d_2)$, in the detector frame (denoted by the superscript ``$d$''). Gravitational-wave detectors are sensitive to detector-frame mass, rather than source-frame mass. This means that the values observed in the detector frame have suffered the effect of the cosmological expansion and are related to their value in the source frame (labelled with the superscript ``$s$'') by the relation $m^s = m^d/(1+z)$. If the mass distribution of compact binaries contains features this breaks the mass-redshift degeneracy, meaning that the observed population of compact objects becomes cosmologically informative. If the attributes of these features are known then they can be strongly informative, with the caveat that assuming an incorrect mass distribution will have a strong impact on the inferred cosmology (see \eg \cite{GWTC-3:cosmology,PhysRevD.104.062009}).

In order to avoid biasing the cosmological inference by assuming a mass distribution, the uncertainty in the population can be marginalised over, by carrying out multi-parameter estimation on cosmological and population hyper-parameters together. Explicitly, the $p(\theta|\Lambda,I)$ terms in Eq. \ref{Eq:sum_likelihood_compact} can be expanded as $p(\theta|\Lambda,I)=p(m^s_{1}, m^s_2|\Lambda,I)p(\phi|\Lambda,I)$, where $\theta=\{m^s_{1}, m^s_2,\phi\}$ and $\Lambda$ is the set of cosmological and population hyper-parameters of interest, including those which describe source-frame mass distribution prior. For the sake of compactness, we do not make this expansion for the rest of the derivation, but it is important to remember that redshift information from the mass distribution is entering the analysis in this way.

In the galaxy catalogue case, instead of taking the redshift prior to be uniform in comoving volume, it is built from a galaxy catalogue, and so instead of being a smooth function of redshift, becomes something highly discrete due to the point-like nature of galaxies.\footnote{Galaxies are, for our purposes, point-like in right ascension and declination. In terms of redshift, however, galaxies often come with a non-negligible redshift uncertainty.} In this scenario, information from the structures observed in the mass distribution still contributes to the cosmological inference as before, though it can be a subdominant effect. \footnote{Information from the mass distribution is likely to be subdominant when the \ac{GW} events used in the analysis are well localised in regions of the sky which have a high level of galaxy catalogue completeness. Conversely, where the localisation volumes of \ac{GW} events are very large, or when the galaxy catalogue is very incomplete, the redshift information from the galaxy catalogue can be subdominant to the information coming from the overall population, and the resulting cosmological inference will converge to the same result as a population-only analysis. }
In addition to galaxies being discrete, galaxy catalogues are incomplete at the redshifts to which \acp{GW} are currently detectable, meaning that an incompleteness correction must be applied to account for the fact that the host galaxy of the \ac{GW} event may not be contained within the catalogue at all, due to being fainter than the flux limit of the galaxy survey. To further complicate matters, the completeness of galaxy catalogues is not uniform across the sky due to 1) obscuration by the Milky Way and 2) different surveys in different patches of the sky, leading to varying depths of completeness as a function of sky location, $\Omega$. The variation of incompleteness across the sky was addressed extensively in \cite{Gray2022}, in which the sky was separated into $N_\text{pix}$ equally-sized pixels and the completeness correction was based on the catalogue's apparent magnitude threshold within each pixel. Here, a similar approach is applied, but by deriving the rest of the method in a different order to the one presented in that paper, the analysis can be made much more computationally tractable. With this in mind, Eq. \ref{Eq:sum_likelihood_compact} can be rewritten to explicitly include the marginalisation over $\Omega$ as a sum over pixels:
\small
\begin{equation}
\begin{aligned}
p(\Lambda|\{x_\text{GW}\},\{D_\text{GW}\},I) 
\propto p(\Lambda|I) p(N_\text{det}|\Lambda,I) &\left[\int \sum^{N_\text{pix}}_j p(D_\text{GW}|\Omega_j,\theta,\Lambda,I) p(\theta|\Omega_j,\Lambda,I) p(\Omega_j|I) d\theta \right]^{-N_\text{det}} \\ &\hspace{-10pt}\times \prod^{N_\text{det}}_i \left[ \int \sum^{N_\text{pix}}_j p(x_{\text{GW}i}|\Omega_j,\theta,\Lambda,I) p(\theta|\Omega_j,\Lambda,I) p(\Omega_j|I) d\theta \right],\\
\propto p(\Lambda|I) p(N_\text{det}|\Lambda,I) &\left[\int \sum^{N_\text{pix}}_j p(D_\text{GW}|\Omega_j,\theta,\Lambda,I)p(\theta|\Omega_j,\Lambda,I) d\theta \right]^{-N_\text{det}} \\ &\times \prod^{N_\text{det}}_i \left[ \int \sum^{N_\text{pix}}_j p(x_{\text{GW}i}|\Omega_j,\theta,\Lambda,I)p(\theta|\Omega_j,\Lambda,I) d\theta \right].
\end{aligned}
\end{equation}
\normalsize
The $p(\Omega_j|I)$ terms are recognised as being identical between same-sized pixels, so come out the front and cancel in numerator and denominator. 

Now it is worth explicitly writing the marginalisation over $z$ (with $\theta = \{z, \theta'\}$):
\small
\begin{equation}\label{Eq:mcmc_framework}
\begin{aligned}
p(\Lambda|\{x_\text{GW}\},\{D_\text{GW}\},I) \hspace{30pt}&\\
\propto p(\Lambda|I) p(N_\text{det}|\Lambda,I) &\left[\iint \sum^{N_\text{pix}}_j p(D_\text{GW}|\Omega_j,z,\theta',\Lambda,I) p(\theta'|\Omega_j,\Lambda,I) p(z|\Omega_j,\Lambda,I) d\theta' dz \right]^{-N_\text{det}} \\ &\times \prod^{N_\text{det}}_i \left[ \iint \sum^{N_\text{pix}}_j p(x_{\text{GW}i}|\Omega_j,z,\theta',\Lambda,I) p(\theta'|\Omega_j,\Lambda,I) p(z|\Omega_j,\Lambda,I) d\theta' dz \right] \\
\propto p(\Lambda|I) p(N_\text{det}|\Lambda,I) &\left[\iint p(D_\text{GW}|z,\theta',\Lambda,I) p(\theta'|\Lambda,I) \sum^{N_\text{pix}}_j  p(z|\Omega_j,\Lambda,I) d\theta' dz \right]^{-N_\text{det}} \\ &\times \prod^{N_\text{det}}_i \left[ \iint \sum^{N_\text{pix}}_j p(x_{\text{GW}i}|\Omega_j,z,\theta',\Lambda,I) p(\theta'|\Lambda,I) p(z|\Omega_j,\Lambda,I) d\theta' dz \right].
\end{aligned}
\end{equation}
\normalsize
Here the assumption has been made that $p(D_\text{GW}|\Lambda,I)$ is uniform across the sky (\ie pixel-independent), and hence can be taken outside the sum over pixels. As $p(D_\text{GW}|\Lambda,I)$ is averaged over an observing run, the rotation of the Earth blurs out much of the \ac{ra} and \ac{dec} dependence of the detectors' antenna patterns, leaving only a mild declination dependence that for the time being can be safely ignored. The remaining parameters contained within $\theta$ are independent of $\Omega$ and so lose their dependence. If the prior on the \ac{GW} merger rate, $R$, is set to $\propto 1/R$, the term $p(N_\text{det}|\Lambda,I)$ loses its dependence on $\Lambda$ and can be ignored (see~\cite{Fishbach:2018edt}), a simplification which is used throughout the rest of this paper.


\subsection{The line of sight redshift prior}

Now the \ac{LOS} redshift prior, $p(z|\Omega_j,\Lambda,I)$, can be considered more closely. As in the original \gwcosmo methodology, the redshift prior must include the cases where the host galaxy is ($G$), and is not ($\bar{G}$), inside the galaxy catalogue, which also requires marginalising over apparent and absolute magnitude ($m$ and $M$) as these determine, to leading order, which galaxies will end up inside a galaxy catalogue which is formed from a flux-limited \ac{EM} survey. An important distinction to make at this point is that between a \emph{galaxy} and a \emph{host galaxy}. So far in the derivation, the assumption that a real \ac{GW} signal has been emitted has been implicit (\ie $x_\text{GW}$ is assumed to correspond to a real \ac{GW} trigger, not a false alarm). This can be made explicit by including the parameter $s$ (denoting the presence of a \ac{GW} source, which until now has been hidden inside $I$ on the right-hand side of every expression). 
\small
\begin{equation}\label{Eq:zprior_start}
\begin{aligned}
p(z|\Omega_i,\Lambda ,s,I) &= \iint \sum_{g=G,\bar{G}} p(z,M,m,g|\Omega_i,\Lambda ,s,I) \; dM dm, \\
&= p(G|\Omega_i,\Lambda,s,I) \iint p(z,M,m|G,\Omega_i,\Lambda,s,I)  \; dM dm \\ & \hspace{100pt}+  p(\bar{G}|\Omega_i,\Lambda,s,I) \iint p(z,M,m|\bar{G},\Omega_i,\Lambda,s,I) \; dM dm,
\end{aligned} 
\end{equation}
\normalsize
\sloppy
As in \cite{Gray2022}, $p(z,M,m|G,\Omega_i,\Lambda,s,I)$ is the prior on $z$, $M$ and $m$ for host galaxies, informed by the galaxy catalogue, within the sky area covered by pixel $i$. Its counterpart, $p(z,M,m|\bar{G},\Omega_i,\Lambda,s,I)$ is the prior on the same parameters but outside the catalogue, and so will have some associated \ac{mth}.

\subsubsection{The in-catalogue part}

Taking a closer look at the in-catalogue part, $p(z,M,m|G,\Omega_i,\Lambda,s,I)$, galaxies come as (approximately) delta-function-like points in $\Omega$ and $m$, with some larger uncertainty on $z$. The absolute magnitude of the galaxy is not provided in the catalogue, and requires a cosmological assumption to compute. For galaxy $j$, $M_j = f(z_j,m_j,\Lambda)$. Applying Bayes' theorem to the term $p(z,M,m|G,\Omega_i,\Lambda,s,I)$ gives:
\small
\begin{equation}
\begin{aligned}
p(z,M,m|G,\Omega_i,\Lambda,s,I) &= \dfrac{p(z,M,m|G,\Omega_i,\Lambda,I)p(s|z,M,m,G,\Omega_i,\Lambda,I)}{p(s|G,\Omega_i,\Lambda,I)},\\
&=  \dfrac{1}{p(s|G,\Omega_i,\Lambda,I)}p(M|z,m,G,\Omega_i,\Lambda,I) p(z,m|G,\Omega_i,\Lambda,I) p(s|z,M,\Lambda,I),\\
&= \dfrac{1}{p(s|G,\Omega_i,\Lambda,I)} \delta(M-M(z,m,\Lambda)) p(z,m|G,\Omega_i,I) p(s|z,M,\Lambda,I).\\
\end{aligned} 
\end{equation}
\normalsize
The value for $M$ is fixed for given values of $z$, $m$ and a cosmological model, and so turns into a delta function. The term $p(z,m|G,\Omega_i,I)$ is simply the prior on $z$ and $m$ given the galaxies in the catalogue within pixel $i$. Note that this is now the prior on \emph{all} galaxies, not just host galaxies, as the parameter $s$ has been separated out. Because this information comes purely from the galaxy catalogue, the dependence on $\Lambda$ is lost. The probability for a galaxy to host a \ac{GW} merger retains some dependence on $z$ and $M$ to allow for possible luminosity weighting, or evolving merger rates with redshift (\ie the $p(s|z,M,\Lambda,I)$ term, which allows galaxies with certain properties to be considered as more likely hosts for \acp{CBC}). Now integrating $p(z,M,m|G,\Omega_i,\Lambda,s,I)$ over $m$ and $M$ leads to 
\small
\begin{equation}\label{Eq:pz_G_beforegals}
\begin{aligned}
\iint p(z,M,m|G,\Omega_i,\Lambda,s,I) \; dM dm &= \dfrac{1}{p(s|G,\Omega_i,\Lambda,I)} \int p(z,m|G,\Omega_i,I) p(s|z,M(z,m,\Lambda),\Lambda,I)\; dm.\\
\end{aligned} 
\end{equation}
\normalsize
The delta-function relationship between $M$, $z$ and $m$ means that the integral over $M$ can be carried out by simply replacing all occurrences of $M$ with $M(z,m,\Lambda)$.

The galaxy catalogue information comes into the term $p(z,m|G,\Omega_i,I)$, which can be replaced by a sum over galaxies. The measured apparent magnitudes (in a given observation band) can be taken as delta functions,\footnote{In reality, these galaxies will have some uncertainty on their measured apparent magnitude which is expected to be small relative to the measured value, and we ignore this as a simplifying assumption (as has previously been done in the literature). An in-depth look into the impact of including this uncertainty is left for future investigation.} while the measured redshift of each galaxy will come with some non-negligible uncertainty, particularly in the case of photometric redshifts. 
\begin{equation}
\begin{aligned}
p(z,m|G,\Omega_i,I) &= \dfrac{1}{N_\text{gal}(\Omega_i)} \sum^{N_\text{gal}(\Omega_i)}_k p(z|\hat{z}_k) \delta(m-\hat{m}_k).
\end{aligned} 
\end{equation}
Here the hat notation is used for measured quantities. The term $p(z|\hat{z}_k)$ is the \emph{posterior} on the redshift of galaxy $k$, given the \emph{measured} redshift $\hat{z}_k$. The rest of this paper will use the assumption that 
\begin{equation}
\begin{aligned}
p(z|\hat{z}_k) = \mathcal{G}(z-\hat{z}_k;\hat{\sigma}_k),
\end{aligned} 
\end{equation}
\ie that a Gaussian centered at $\hat{z}_k$ with standard deviation $\hat{\sigma}_k$ is a reasonable approximation for the redshift posterior of galaxy $k$.\footnote{While this implementation was chosen for ease, there is no strict requirement that the galaxy uncertainties be Gaussian, or even that they follow the same distribution. Hypothetically this method would very easily extend to include the full PDF for photometric redshifts, allowing their often non-Gaussian structure to be captured.} Additionally the assumption is made that the galaxy measurements provided  in the catalogue used for the analyses later in this paper (see Sec. \ref{sec:LOS_practical}) are posteriors.\footnote{If the measured redshift is actually a likelihood, a prior ought to be applied, in which case a uniform in comoving volume prior might be sensible, to account for the larger volume at higher redshifts. In the limit that $\hat{\sigma}_k \rightarrow 0$, and $p(z|\hat{z}_k)$ becomes a delta function, the choice of prior will become irrelevant. This is yet another reason to look forward to spectroscopic redshift measurements.}

Returning to Eq. \ref{Eq:pz_G_beforegals}, and substituting the sum over galaxies into the integral over $m$ returns the following:
\small
\begin{equation}\label{Eq:pz_G}
\begin{aligned}
\iint p(z,M,m|G,\Omega_i,\Lambda,s,I) \; dM dm &= \dfrac{1}{p(s|G,\Omega_i,\Lambda,I)} \dfrac{1}{N_\text{gal}(\Omega_i)} \\ & \hspace{30pt}\times \int \sum^{N_\text{gal}(\Omega_i)}_k p(z|\hat{z}_k) \delta(m-\hat{m}_k) p(s|z,M(z,m,\Lambda),\Lambda,I) \; dm,\\
&= \dfrac{1}{p(s|G,\Omega_i,\Lambda,I)} \dfrac{1}{N_\text{gal}(\Omega_i)}\sum^{N_\text{gal}(\Omega_i)}_k p(z|\hat{z}_k) p(s|z,M(z,\hat{m}_k,\Lambda),\Lambda,I).
\end{aligned} 
\end{equation}
\normalsize
This in-catalogue part of the \ac{LOS} redshift prior is essentially a weighted sum of Gaussians, where the weights are some function of redshift and absolute magnitude.

\subsubsection{The out-of-catalogue part}

Now turning attention to the out-of-catalogue part, $p(z,M,m|\bar{G},\Omega_i,\Lambda,I)$, and expanding it leads to a slightly different expression:
\small
\begin{equation}\label{Eq:outofcatprior_start}
\begin{aligned}
p(z,M,m|\bar{G},\Omega_i,\Lambda,s,I) &= \dfrac{p(z,M,m|\bar{G}, \Omega_i,\Lambda,I) p(s|z,M,m,\bar{G},\Omega_i,\Lambda,I)}{p(s|\bar{G},\Omega_i,\Lambda,I)}, \\
&= \dfrac{1}{p(s|\bar{G},\Omega_i,\Lambda,I)} p(z,M,m|\bar{G},\Omega_i,\Lambda,I) p(s|z,M,\Lambda,I), \\
&= \dfrac{1}{p(s|\bar{G},\Omega_i,\Lambda,I)} \dfrac{p(\bar{G}|z,M,m,\Omega_i,\Lambda,I)p(z,M,m|\Omega_i,\Lambda,I)}{p(\bar{G}|\Omega_i,\Lambda,I)} p(s|z,M,\Lambda,I).
\end{aligned} 
\end{equation}
\normalsize
Here an additional step has been taken to separate out the term $p(\bar{G}|z,M,m,\Omega_i,\Lambda,I)$, which is the probability that a galaxy with properties $z$, $M$, $m$, etc. is outside the galaxy catalogue, which requires modeling the \ac{EM} selection effects of that particular survey. The main assumption here is that the galaxy catalogue is flux limited, such that the probability that a galaxy is or is not contained within it depends on the galaxy's apparent magnitude, and whether it is greater or less than the apparent magnitude threshold of the catalogue along the same line of sight, $m_\text{th}(\Omega_i)$. However, this is not the only reason galaxies may be missing from a particular catalogue. If the redshift or colour information of galaxies at high redshifts becomes particularly unreliable then it may be prudent to discard galaxies above some redshift, $z_\text{cut}$, from the catalogue, in which case their exclusion must be accounted for with the out of catalogue term. In this case it becomes
\begin{equation}\label{Eq:Gbar_mth_z}
\begin{aligned}
p(\bar{G}|z,M,m,\Omega_i,H_0,I) &= \Theta[m-m_\text{th}(\Omega_i)]\Theta[z_\text{cut}-z] + \Theta[z-z_\text{cut}].
\end{aligned} 
\end{equation}
Here, $\Theta$ denotes a Heaviside step function. If no redshift information is discarded, \ie $z_\text{cut} \rightarrow \infty$, then $\Theta[z_\text{cut}-z] =1$ and  $\Theta[z-z_\text{cut}] = 0$, meaning that the selection effects simplify to only depending on the \ac{mth} of the catalogue.

Another term to examine more closely is $p(z,M,m|\Omega_i,\Lambda,I)$, which is the prior on redshift, absolute and apparent magnitude for (all) galaxies within pixel $i$. Because the parameter $\bar{G}$ is not present here, this term does not depend on direction-dependent \ac{EM} selection effects, and so $\Omega_i$ can be dropped from the right-hand side. Due to the relationship between $z$, $M$ and $m$, priors only need to be defined for two of these parameters, as the third is specified by its relationship to the other two. For the out of catalogue case it makes sense to define priors for $z$ and $M$ because these are reasonably well known (galaxies are broadly expected to follow a uniform in comoving volume distribution in redshift when viewed on a large enough scale, and galaxy absolute magnitudes are expected to follow a Schechter function). This means that the term can be rewritten as $p(z,M,m|\Lambda,I) = \delta(m-m(z,M,\Lambda)) p(z,M|\Lambda,I)$. Substituting this and Eq. \ref{Eq:Gbar_mth_z} into Eq. \ref{Eq:outofcatprior_start} gives 
\small
\begin{equation}\label{Eq:outofcatprior_end}
\begin{aligned}
p(z,M,m|\bar{G},\Omega_i,\Lambda,s,I) &= \dfrac{1}{p(s|\bar{G},\Omega_i,\Lambda,I) p(\bar{G}|\Omega_i,\Lambda,I)} \left(\Theta[m-m_\text{th}(\Omega_i)]\Theta[z_\text{cut}-z] + \Theta[z-z_\text{cut}]\right) \\ &\hspace{100pt}\times  \delta(m-m(z,M,\Lambda)) p(z,M|\Lambda,I) p(s|z,M,\Lambda,I).
\end{aligned} 
\end{equation}
\normalsize

Now integrating this term over both $m$ and $M$ leads to
\small
\begin{equation}\label{Eq:pz_Gbar}
\begin{aligned}
\iint &p(z,M,m|\bar{G},\Omega_i,\Lambda,s,I) \; dM dm \\&= \dfrac{1}{p(s|\bar{G},\Omega_i,\Lambda,I) p(\bar{G}|\Omega_i,\Lambda,I)} \int \left( \Theta[m(z,M,\Lambda)-m_\text{th}(\Omega_i)] \Theta[z_\text{cut}-z] + \Theta[z-z_\text{cut}] \right) \\ &\hspace{200pt}\times p(z,M|\Lambda,I) p(s|z,M,\Lambda,I) \; dM, \\
&= \dfrac{1}{p(s|\bar{G},\Omega_i,\Lambda,I) p(\bar{G}|\Omega_i,\Lambda,I)} \left[\Theta[z_\text{cut}-z] \int^{M_\text{max}(H_0)}_{M(z,m_\text{th}(\Omega_i),\Lambda)} p(z,M|\Lambda,I) p(s|z,M,\Lambda,I) \; dM \right. \\& \hspace{150pt}+ \left. \Theta[z-z_\text{cut}]  \int^{M_\text{max}(H_0)}_{M_\text{min}(H_0)} p(z,M|\Lambda,I) p(s|z,M,\Lambda,I) \; dM\right]. 
\end{aligned} 
\end{equation}
\normalsize
The Heaviside step function for the apparent magnitude threshold has been converted to a redshift-dependent integration limit over $M$ which depends on the apparent magnitude threshold of pixel $i$. The integration limits on $M$ in general depend on $H_0$ because the parameters of the Schechter function are \ac{H0}-dependent, but the final distribution is insensitive to its value.

\subsubsection{The full expression for the LOS prior}
Both Eq. \ref{Eq:pz_G} and Eq. \ref{Eq:pz_Gbar} contain prefactors which are of a format hard to compute. They take the form $1/p(s|G,\Omega_i,\Lambda,I)$ for the former, and $1/(p(s|\bar{G},\Omega_i,\Lambda,I) p(\bar{G}|\Omega_i,\Lambda,I))$ for the latter. However, when substituting Eq. \ref{Eq:pz_G} and Eq. \ref{Eq:pz_Gbar} into Eq. \ref{Eq:zprior_start} it can be seen that these expressions simplify. For ease of notation let the rest of Eq. \ref{Eq:pz_G} and Eq. \ref{Eq:pz_Gbar} be termed $f(\text{IN CAT})$ and $f(\text{OUT OF CAT})$ respectively. Then the expression becomes
\small
\begin{equation}\label{Eq:prefactors}
\begin{aligned}
p(z|\Omega_i,\Lambda,s,I) &= \dfrac{p(G|\Omega_i,\Lambda,s,I)}{p(s|G,\Omega_i,\Lambda,I)} f(\text{IN CAT})  + \dfrac{p(\bar{G}|\Omega_i,\Lambda,s,I)}{p(s|\bar{G},\Omega_i,\Lambda,I) p(\bar{G}|\Omega_i,\Lambda,I)} f(\text{OUT OF CAT}), \\
&= \dfrac{p(s|G,\Omega_i,\Lambda,I)p(G|\Omega_i,\Lambda,I)}{p(s|G,\Omega_i,\Lambda,I)p(s|\Omega_i,\Lambda,I)}  f(\text{IN CAT}) \\& \hspace{100pt} + \dfrac{p(s|\bar{G},\Omega_i,\Lambda,I)p(\bar{G}|\Omega_i,\Lambda,I)}{p(s|\Omega_i,\Lambda,I)p(s|\bar{G},\Omega_i,\Lambda,I) p(\bar{G}|\Omega_i,\Lambda,I)} f(\text{OUT OF CAT}), \\
&= \dfrac{1}{p(s|\Omega_i,\Lambda,I)} \left[p(G|\Omega_i,\Lambda,I)  f(\text{IN CAT}) + f(\text{OUT OF CAT}) \right].
\end{aligned} 
\end{equation}
\normalsize
There is a lack of symmetry between the in and out of catalogue terms, as one might have expected a $p(\bar{G}|\Omega_i,H_0,I)$ as a prefactor to the out of catalogue term, but this term has cancelled with one in the denominator. With further consideration, this makes sense -- the sum over galaxies of the in-catalogue part by definition integrates to 1 over $z$ (assuming uniform galaxy weights for now), while the integral over the out-of-catalogue part does not, due to the apparent magnitude threshold appearing in the integral limits.

The term $1/p(s|\Omega_i,\Lambda,I)$ can be ignored as a normalisation constant. It will be the same for all pixels, assuming \ac{GW} sources are isotropically distributed, and as such will eventually cancel with an identical term in denominator once the \ac{LOS} prior is substituted back into Eq. \ref{Eq:mcmc_framework}.

The one term we have not yet looked at is the probability of a galaxy in pixel $i$ being inside the galaxy catalogue (notice now that there is no $s$ on the right-hand side, so here we are considering \emph{all} galaxies, not just host ones):
\begin{equation}\label{Eq:pGstart}
\begin{aligned}
p(G|\Omega_i,\Lambda,I) &= \iiint p(G|z,M,m,\Omega_i,\Lambda,I) p(z,M,m|\Omega_i,\Lambda,I) \; dz dM dm.
\end{aligned} 
\end{equation}
The probability that a galaxy is inside the catalogue can then be expanded as
\begin{equation}
\begin{aligned}
p(G|z,M,m,\Omega_i,\Lambda,I) &= \Theta[m_\text{th}(\Omega_i)-m]\Theta[z_\text{cut}-z],
\end{aligned} 
\end{equation}
which leads to
\begin{equation}\label{Eq:pGend}
\begin{aligned}
p(G|\Omega_i,\Lambda,I) &= \int^{z_\text{cut}}_0 \int^{M(z,m_\text{th}(\Omega_i),\Lambda)}_{M_\text{min}(H_0)} p(z,M|\Lambda,I) \; dM dz. \\
\end{aligned} 
\end{equation}

The fact that all probability densities must be properly normalised may cause the reader to wonder whether the choice of redshift range over which this is done so has an impact on this analysis. After all the term $p(G|\Omega_i,\Lambda,I)$ will take a very different value if $p(z,M|\Lambda,I)$ is normalised over a redshift range which is truncated at $z_\text{max}=2$ and one where $z_\text{max}=10$. However, it is useful to note that as long as the same expression for $p(z,M|\Lambda,I)$ is also be used in Eq. \ref{Eq:pz_Gbar}, the normalisation will come out the front as a constant. As such, the choice of what redshift range over which to normalise $p(z,M|\Lambda,I)$ has no bearing on the result.

Combining Eqs. \ref{Eq:prefactors}, \ref{Eq:pz_G} and \ref{Eq:pz_Gbar} leads to a full expression for the redshift prior of
\small
\begin{equation}\label{Eq:pz_withgals}
\begin{aligned}
p(z|\Omega_i,\Lambda,s,I) = \dfrac{1}{p(s|\Omega_i,\Lambda,I)} \Bigg[ &p(G|\Omega_i,\Lambda,I) \dfrac{1}{N_\text{gal}(\Omega_i)}\sum^{N_\text{gal}(\Omega_i)}_k p(z|\hat{z}_k) p(s|z,M(z,\hat{m}_k,\Lambda),\Lambda,I) \\ &+ \bigg(\Theta[z_\text{cut}-z] \int^{M_\text{max}(H_0)}_{M(z,m_\text{th}(\Omega_i),\Lambda)} p(z,M|\Lambda,I) p(s|z,M,\Lambda,I) \; dM  \\& \hspace{10pt}+ \Theta[z-z_\text{cut}]  \int^{M_\text{max}(H_0)}_{M_\text{min}(H_0)} p(z,M|\Lambda,I) p(s|z,M,\Lambda,I) \; dM\bigg) \Bigg].
\end{aligned} 
\end{equation}
\normalsize
Looking ahead, Fig. \ref{fig: LOS zprior examples} shows an example of the \ac{LOS} redshift prior, demonstrating the discrete galaxies which are visible at low redshift, and how it converges to a uniform in comoving distribution at higher redshifts as galaxy catalogue completeness decreases.

\subsubsection{The separability of GW mass and rate evolution hyper-parameters from the line-of-sight redshift prior \label{sec:rate_separation}}

The \ac{LOS} redshift prior, as presented in Eq. \ref{Eq:pz_withgals}, retains clear dependence on $\Lambda$ throughout, where $\Lambda$ is a set of cosmological and population hyper-parameters of interest, many of which we would like to jointly estimate. However, if Eq. \ref{Eq:pz_withgals} has to be recomputed for every newly drawn set of parameter values this will slow things down to the point of impracticality. To address this, $\Lambda$ can be broken into its constituent parts, namely $\Lambda = \{\Lambda_\text{cosmo}, \Lambda_\text{mass}, \Lambda_\text{rate}\}$.  $\Lambda_\text{cosmo}$ denotes the parameters of the cosmological model, namely $\{H_0, \Omega_M, \Omega_\Lambda\}$. $\Lambda_\text{mass}$ denotes the parameters of the \ac{GW} mass model, such as minimum and maximum mass, slope, the position of any features, etc. $\Lambda_\text{rate}$ denotes the parameters of the rate-evolution model which determines how the intrinsic rate of \ac{CBC} mergers evolves with redshift (see Sec. \ref{sec:results} for details).

The immediate impact of substituting this definition into Eq. \ref{Eq:pz_withgals} is that the priors on galaxy redshift and absolute magnitude depend only on $\Lambda_\text{cosmo}$. Beyond this, some simplifying assumptions must be made. We assume that the \ac{GW} mass model doesn't evolve with redshift (there is, as yet, no strong evidence for a mass model which evolves with redshift \cite{PhysRevX.13.011048}, though this may change once the data from future observing runs has been analysed  \cite{2022MNRAS.515.5495M,2023MNRAS.523.4539K}). In this scenario, none of the terms in Eq. \ref{Eq:pz_withgals} are dependent on $\Lambda_\text{mass}$. The host galaxy weights term retains dependence on $\Lambda_\text{rate}$ (and, in the case where those galaxies come from the catalogue, such that their absolute magnitude is computed from their apparent magnitude and redshift, $\Lambda_\text{cosmo}$).

So far in Eq. \ref{Eq:pz_withgals} the only assumption about the relationship between \ac{GW} mergers and host galaxies is that it will have some (unspecified) dependence on redshift and absolute magnitude. At this point in time, the true relationship between galaxy properties and \ac{GW} merger probability is not well constrained, but there are a couple of reasons for parameterising the problem this way. The usual justification is to reason that there is likely a causal link between galaxy mass and/or \ac{SFR} and that galaxy's probability of hosting a \ac{GW} merger. Higher mass means more matter available to be turned into compact binaries. Higher \ac{SFR} means more stars forming, which means more binary systems forming and eventually turning into compact binaries. Of course there may well be some significant time-delay between \ac{SFR} and a binary merging, but this is also currently uncertain. In any case, both mass and \ac{SFR} are traced by galaxy luminosity in different bands, and so estimates for both can be constructed from observable data.\footnote{Of course other things such as galaxy metallicity and galaxy type may all also play a role in host galaxy probability, but a full exploration of these parameters and how they could be incorporated into the current methodology is left for the future.}

As we do not yet know the exact relationship between \ac{GW} mergers and host galaxy properties, it is worth remaining flexible and allowing that there may be some additional redshift-dependence to the merger rate model. As such, for the remainder of this paper, the following form is taken:
\begin{equation}
\begin{aligned}
p(s|z,M,\Lambda_\text{rate},I) &\propto p(s|z,\Lambda_\text{rate},I) p(s|M,I).
\end{aligned} 
\end{equation}
where
\begin{equation}
p(s|M,I) \propto
\begin{cases}
 L(M)  &\text{if using luminosity weighting} \\
 \text{const} &\text{if assuming uniform weighting.}
\end{cases} 
\end{equation}
By keeping the parameter(s) $\Lambda_\text{rate}$ flexible, such that $p(s|z,\Lambda_\text{rate},I)$ can return to a uniform distribution, this allows additional unknown redshift dependence to be captured.

Making this substitution in Eq. \ref{Eq:pz_withgals} leads to
\small
\begin{equation}\label{Eq:pz_withgals_sep}
\begin{aligned}
p(z|\Omega_i,\Lambda,s,I) = &\dfrac{p(s|z,\Lambda_\text{rate},I)}{p(s|\Omega_i,\Lambda,I)} \Bigg[ p(G|\Omega_i,\Lambda,I) \dfrac{1}{N_\text{gal}(\Omega_i)}\sum^{N_\text{gal}(\Omega_i)}_k p(z|\hat{z}_k) p(s|M(z,\hat{m}_k,\Lambda_\text{cosmo}),I) \\ &+ p(z|\Lambda_\text{cosmo},I) \bigg( \Theta[z_\text{cut}-z] \int^{M_\text{max}(H_0)}_{M(z,m_\text{th}(\Omega_i),\Lambda_\text{cosmo})} p(M|z,\Lambda_\text{cosmo},I) p(s|M,I) \; dM  \\ &\hspace{100pt} + \Theta[z-z_\text{cut}]  \int^{M_\text{max}(H_0)}_{M_\text{min}(H_0)}  p(M|z,\Lambda_\text{cosmo},I) p(s|M,I) \; dM \bigg) \Bigg].
\end{aligned} 
\end{equation}
\normalsize
The term $p(s|z,\Lambda_\text{rate},I)$ has come out the front, meaning that $\Lambda_\text{rate}$ parameters are not needed for computing the bulk of the \ac{LOS} prior, and the aim becomes to simply precompute the part of the expression within the square brackets.

Within the square brackets there is still a clear dependence on $\Lambda_\text{cosmo}$ in both the in- and out-of-catalogue parts. The term $p(z|\Lambda_\text{cosmo},I)$ is a uniform in comoving volume distribution and its dependence on \ac{H0} drops out when normalised, leaving only the dependence on $\{\Omega_M, \Omega_\Lambda\}$. The galaxy luminosity function, $p(M|z,\Lambda_\text{cosmo},I)$, will depend only on \ac{H0} (Schechter function parameters are quoted with a dependence on $h \equiv H_0/100$). The relation between redshift, apparent and absolute magnitude, $M(z,m,\Lambda_\text{cosmo})$ will retain dependence on all cosmological parameters due to the necessity of converting redshift to luminosity distance.

Despite the dependence of individual terms on $H_0$, when computed for different values of $H_0$ the \ac{LOS} prior is independent of its value down to some normalisation constant. Because the same prior is used to evaluate the \ac{GW} likelihood and \ac{GW} selection effects, this dependence cancels, meaning that the \ac{LOS} prior can in fact be computed just once and used in an analysis in which $\{H_0, \Lambda_\text{mass}, \Lambda_\text{rate}\}$ vary.

\subsubsection{The resolution of the LOS prior}
\label{sec: The resolution of the LOS prior}

Looking at Eq. \ref{Eq:pz_withgals_sep} it should be noted that pixel area does not explicitly appear anywhere within the current definition, but there is an implicit assumption present which is worth taking care with. The quantities computed in the method thus far are based on assumptions about the distribution of galaxies which we cannot see (\ie that they are distributed, on average, uniformly in comoving volume, and follow some kind of luminosity distribution such as a Schechter function). Consider $p(G|\Omega_i,\Lambda,I)$ for example, and how it is expressed in Eq. \ref{Eq:pGend}. This term is the fraction of galaxies which within pixel $i$ which are inside the galaxy catalogue. The ``true'' value is given by $N_\text{gal}(\Omega_i)/N_\text{tot}(\Omega_i)$ -- that is the ratio of the number of galaxies in the catalogue (in that pixel) to the total number of galaxies (including the ones we cannot see) which lie in that pixel. Because we do not have access to the true total number of galaxies we must use an approximation based on our chosen priors. This approximation is okay, as long as the pixel size is large enough that $N_\text{gal}(\Omega_i) \approx p(G|\Omega_i,\Lambda,I) \times N_\text{tot} \times \Delta \Omega$, where $N_\text{tot}$ is the total number of galaxies in the universe, and $\Delta \Omega$ is the fractional area covered by one pixel. This will clearly break down in the limit of infinitesimally small pixels which may not contain any galaxies, but which would still have a non-zero value of $p(G|\Omega_i,\Lambda,I)$. In order to avoid this limit, pixel size must be chosen to be large enough that the approximation is valid.\footnote{Alternatively, if pixels with the same level of completeness could be grouped together, such that $N_\text{gal}(\Omega_i)$ can be estimated over a larger patch of the sky, this issue with small number statistics may be similarly avoided. To do so is beyond the scope of this paper and is left for future work.}

This fine-resolution limit to the method has a drawback, in that all galaxies within a pixel are treated as having the same $\Omega$, effectively reducing the galaxies' positional (\ac{ra} and \ac{dec}) information. If the \ac{GW} likelihood changes significantly across the area of one pixel, valuable information will be lost. In order to carry out the analysis with suitably high resolution, we make the choice to define two different levels of resolution: $n_\text{map}$, which determines the resolution on which $N_\text{gal}(\Omega_i)$ and \ac{mth} are estimated,\footnote{Here we continue to rely on estimating $m_\text{th}(\Omega_i)$ using the median $m$ of the galaxies inside pixel $i$, an assumption which also breaks down in the high-resolution limit.} and $n_\text{high}$, a finer resolution which pixels are subdivided into. The specific choices for these resolutions are discussed in section \ref{sec:LOS_practical}. These parameters, $n_\text{map}$ and $n_\text{high}$ are values of nside, a parameter which relates to the number of equal-sized pixels the surface of a sphere is separated into, using the formula $n_\text{pix} = 12 \times \text{nside}^2$.

The $n_\text{map}$ resolution is used to compute \textsc{HEALPix} maps \cite{2005ApJ...622..759G,Zonca2019} of two quantities:\footnote{\textsc{HEALPix} is available at \url{http://healpix.sf.net}.} \ac{mth} and $N_\text{gal}(\Omega_i)$. When computing the full \ac{LOS} redshift prior on the $n_\text{high}$ resolution, \ac{mth} is taken from the \ac{mth} map, and used for the incompleteness corrections. $N_\text{gal}(\Omega_i)$ is also taken from the coarser map, and divided through by the number of sub-pixels that the coarse pixel is divided into, and this value is then used in place of $N_\text{gal}(\Omega_i)$ in Eq. \ref{Eq:pz_withgals_sep}. For clarity, the sum is still carried out over the true number of galaxies in the fine pixel, but the normalising factor of $1/N_\text{gal}(\Omega_i)$ is constructed from the coarse map. This means that you will get the same results when computing the full \ac{LOS} prior on a coarse map as when computing it on a higher resolution, then summing pixels to return to the same resolution as the coarse map. As such it is robust in the limit of infinitely high resolutions. 

\begin{figure}
    \centering
    \includegraphics[width=0.8\textwidth]{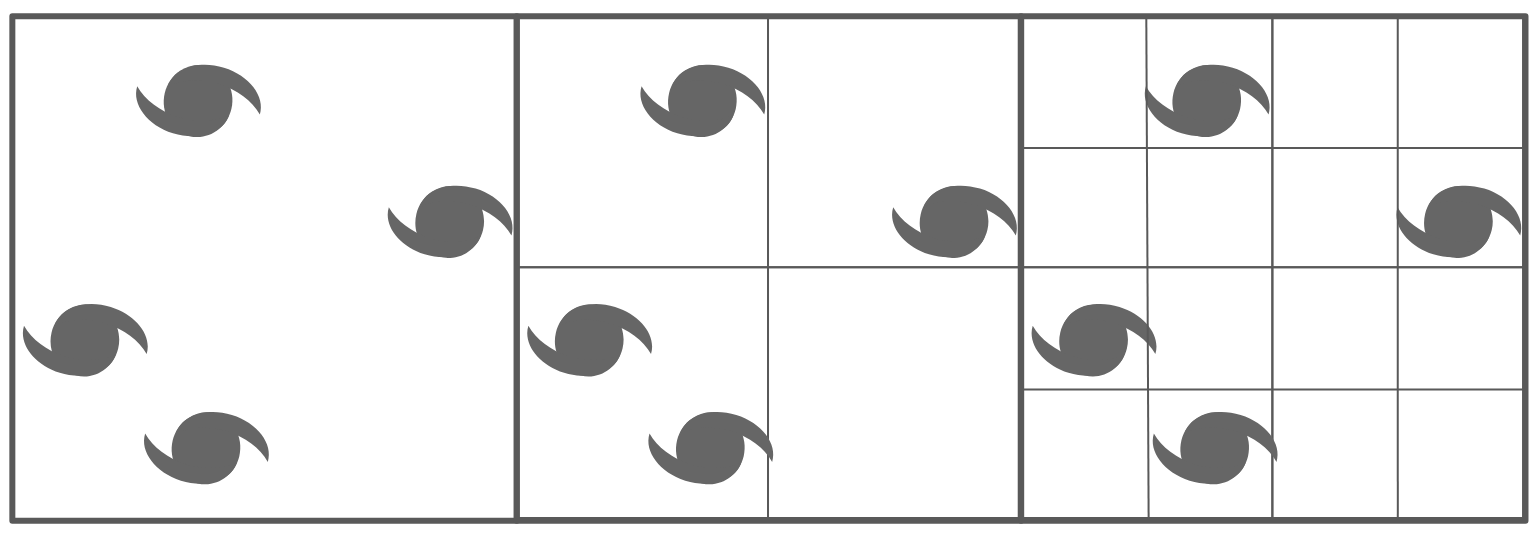}
    \caption{Schematic demonstrating the impacts of pixel resolution. The same patch of sky containing four galaxies is shown with 3 different pixel resolutions.}
    \label{fig:pixel_resolution}
\end{figure}
Figure \ref{fig:pixel_resolution} may help to visualise this. First assume you do not pre-compute \ac{mth} or $N_\text{gal}(\Omega_i)$ on a coarser resolution. Looking at the left-hand panel, if the \ac{LOS} prior was constructed each galaxy would have a normalising factor of $1/4$ in front of it. In the middle panel, however, the pixel has been split into 4 sub-pixels, and the same galaxies have normalisation factors of 1 or $1/2$, depending on which pixel they fall in. If you then summed the 4 sub-pixels you would not recover the same result as the left-hand panel; instead two of your galaxies would be down-weighted relative to the other two, simply due to small number statistics. Additionally, in the high-resolution limit (right-hand panel) there will be a maximum of 1 galaxy per pixel, but there can be an arbitrarily high number of empty pixels, which means the out-of-catalogue contribution will dominate. 
If, however, $N_\text{gal}(\Omega_i)$ is computed for a coarse pixel then divided through by the number of sub-pixels (leading to an average of 4 galaxies per sub-pixel in the left-hand panel of Fig. \ref{fig:pixel_resolution}, 1 for the middle panel, and 1/4 for the right-hand panel), then the effective $N_\text{gal}(\Omega_i)$ of each high resolution pixel remains inversely proportional to the total number of out-of-catalogue contributions (1 per sub-pixel).


\subsection{Gravitational-wave selection effects \label{sec:injections}}

In order to obtain an unbiased estimate of cosmological (and population) parameters, it is crucial to accurately model \ac{GW} selection effects. The ``probability of detection'' of a \ac{GW} event (often shortened to $P_\text{det}$) can be written as $p(D_\text{GW}|\Lambda,I) = \int p(D_\text{GW}|\theta,\Lambda,I)p(\theta|\Lambda,I) \mathrm{d}\theta$. In previous versions of \gwcosmo, where the only parameter being constrained was \ac{H0} (\ie $\Lambda=\Lambda_\text{cosmo}=H_0$), and all other mass model and cosmological parameters were set to fixed values, this was done by simulating a population of \ac{GW} events and computing their probability of detection as a function of redshift and \ac{H0} (marginalising over masses, sky position, polarisation, and inclination). Now, with a dozen or more parameters to jointly constrain, this approach is no longer practical due to the high computational burden.

In order for Eq. \ref{Eq:mcmc_framework} to be used by a Markov Chain Monte Carlo (MCMC) or Nested Sampling method, the probability of detection must be quickly computable for any values of $\Lambda=\{\Lambda_\text{cosmo},\Lambda_\text{mass},\Lambda_\text{rate}\}$. 
Probability of detection is equivalent to the ratio of detectable events $N_\text{exp}(\Lambda)$ over the total number of mergers $N(\Lambda)$ up to a certain redshift. This integral can be computed numerically using a Monte Carlo approximation. 
\begin{equation}
\begin{aligned}
\frac{N_\text{exp}}{N}(\Lambda) &= \iiint p(D_\text{GW}|m_1^s,m_2^s,z,\Lambda,I)p(m_1^s,m_2^s,z|\Lambda,I) \mathrm{d}m_1^s \mathrm{d}m_2^s \mathrm{d}z,\\
 &\approx\frac{1}{N_\text{sim}}\sum_{i=1}^{N_\text{det}}\frac{p(m^s_{1,i},m^s_{2,i},z_{i}|\Lambda,I)}{\pi_\text{inj}^s(m^s_{1,i},m^s_{2,i},z_{i}|\Lambda,I)},\label{eq:importance_sampling}
\end{aligned}
 \end{equation}
 where $\pi_\text{inj}^s$ is used to denote a (prior) probability density function which is distinct from another prior on the same parameters (see later in this section). 
The discrete summation is done over a large set of simulated GW signals (typically of the order of $N_\text{sim}=10^5-10^6$) using a prior on the characteristics of the simulated CBC (masses, redshift or luminosity distance). These signals are injected in realistic time sequences of noise following the detector sensitivities and duty cycles in order to reproduce observational conditions as closely as possible. Then, we can compute the corresponding \ac{SNR} for each simulated signal so that we have in the end a number $N_\text{det}$ of detected events (injections) above the specified network \ac{SNR} threshold in the analysis, which, when summed produces the probability of detection.
 
 The probabilities appearing in the numerator of Eq. \ref{eq:importance_sampling},  $p(m^s_{1,i},m^s_{2,i},z_{i}|\Lambda,I)$, are computed for each value of $\Lambda$ drawn during the sampling process, applying the desired priors on compact binary source-frame masses and redshift (as an example, the prior for the distribution of the primary mass $m_1$ in the source frame can be seen in~Eq.~\ref{eq:powerlaw+peak}). In particular, the model used for $p(z|\Lambda,I)$ is the LOS redshift prior of Eq.~\ref{Eq:pz_withgals_sep}. The probabilities of the denominator, $\pi_\text{inj}^s(m^s_{1,i},m^s_{2,i},z_{i}|\Lambda)$, are the individual probabilities of the detected injections after taking into account the current value of $\Lambda$ and must be recomputed accordingly. These probabilities are therefore closely related to their known values, when the injections have been drawn from the chosen prior distribution $\pi_\text{inj}$. In the following analyses, we chose to draw injections in the detector frame, $\pi_\text{inj}(m^d_{1,i},m^d_{2,i},d_{\mathrm{L},i})$, as detectability of a \ac{GW} event is a function of these parameters, and they therefore provide coverage for a wide range of cosmological values. With this choice, it is possible to obtain the required values in terms of source frame masses, $\pi_\text{inj}^s(m^s_{1,i},m^s_{2,i},z_{i}|\Lambda)$, using the relation:
\begin{equation}
\pi_\text{inj}^s(m^s_{1,i},m^s_{2,i},z_{i}|\Lambda) = \pi_\text{inj}(m^d_{1,i},m^d_{2,i},d_{\mathrm{L},i})(1+z_i)^2\left|\frac{\partial d_L}{\partial z}\right| _{z=z_i}
\label{Eq:jacobians}
\end{equation}
where the Jacobians for transforming between source frame and detector frame, and $d_L$ and $z$ have been written explicitly. Finally, substituting Eq. \ref{Eq:jacobians} into the denominator of Eq. \ref{eq:importance_sampling} gives:
\begin{equation}
\frac{N_\text{exp}}{N}(\Lambda) \approx\frac{1}{N_\text{sim}}\sum_{i=1}^{N_\text{det}}
 \frac{p(m^s_{1,i},m^s_{2,i},z_{i}|\Lambda)}{\pi_\text{inj}(m^d_{1,i},m^d_{2,i},d_{\mathrm{L},i})(1+z_i)^2\left|\frac{\partial d_L}{\partial z}\right| _{z=z_i}}.
 \label{Eq:inj_final}
 \end{equation}
The denominator acts as a weight of each detected injection. The sampling process modifies these weights for each new value of $\Lambda$; this is a fast operation and we are therefore able to get, at every point of the hyper-parameters space, the estimate of $N_\text{exp}/N$ for any value of $\Lambda$. Assuming that the redshift prior in the numerator of Eq. \ref{Eq:inj_final} is constructed from the \ac{LOS} redshift prior, this is (to within a normalisation constant) the quantity between the first set of square brackets appearing in Eq.~\ref{Eq:mcmc_framework}:
\begin{equation*}
\left[\iint p(D_\text{GW}|z,\theta',\Lambda,I) p(\theta'|\Lambda,I) \sum^{N_\text{pix}}_j  p(z|\Omega_j,\Lambda,I) d\theta' dz \right].
\end{equation*}


\section{Gravitational-wave data and galaxy catalogues\label{sec:data}}

\subsection{Gravitational-wave data \label{sec:GWdata}}
In this paper we select all \ac{GW} events with \ac{SNR} $> 11$ from GWTC-3 (as was done in \cite{GWTC-3:cosmology}). This leads to a selection of 42 \acp{BBH}, 2 \acp{NSBH} (GW200105 and GW200115), the asymmetric mass binary GW190814 (which is treated as an \ac{NSBH} in all the analyses which follow) and as well as two \acp{BNS} (GW170817 and GW190425). \gwcosmo makes use of both \ac{GW} posterior samples (for computing a \ac{KDE} of the \ac{GW} distance distribution along every line-of-sight) and skymaps (for weighting the normalised distance \acp{KDE} to retain the correct 3-dimensional distribution of \ac{GW} probability in \ac{ra}, \ac{dec} and luminosity distance).\footnote{The data, in the form of posterior samples and skymaps, for O1, O2 and O3a \cite{2021arXiv210801045T} can be found at \url{https://zenodo.org/record/6513631} , and for O3b \cite{GWTC-3} at \url{https://zenodo.org/record/5546663}.}

A set of injections (as described in section \ref{sec:injections}) was generated using \gwcosmo to characterise the detectability of \ac{GW} events assuming a network \ac{SNR} threshold of 11.

\subsection{Galaxy catalogues \label{sec:LOS_practical}}

For this analysis we use the GLADE+ galaxy catalogue, which is a composite catalogue made by cross-matching the Gravitational Wave Galaxy Catalogue \cite{White:2011qf}, HyperLEDA \cite{Makarov:2014txa}, the 2 Micron All-Sky Survey Extended Source Catalog \cite{2000AJ....119.2498J,Skrutskie:2006wh}, the 2MASS Photometric Redshift Catalog \cite{Bilicki:2013sza}, the WISExSCOS Photometric Redshift Catalogue \cite{Bilicki_2016}, and the Sloan Digital
Sky Survey quasar catalogue from the 16th data release \cite{Lyke_2020}. We choose galaxies in the $K$-band in order for our results to be comparable to \cite{GWTC-3:cosmology}. We assume that the galaxy luminosity function follows Schechter function that does not evolve with redshift,\footnote{This is a reasonable assumption at low redshifts, but will not remain true at higher redshifts ($z \sim 1$). The flexibility in the model which evolves \ac{GW} merger rate with redshift in \gwcosmo will compensate for some mis-match between the chosen Schechter function here and the truth, but this is a point which it will be worth investigating thoroughly in the future.} with parameters $M^* = -23.39 + 5\log(h)$ (where $h \equiv H_0/100$) and $\alpha = -1.09$, following \cite{Kochanek_2001}. To curtail the faint-end of the Schechter function we choose a limit of $M_\text{max} = -19.0 + 5\log(h)$. The default assumption for host galaxy weighting is that host probability is proportional to $K$-band luminosity, which traces galaxy stellar mass. As mentioned in section \ref{sec:rate_separation}, an assumption of \ac{CBC} merger rate evolution with redshift is not required at this stage as it is incorporated at a later stage in the analysis. $K$-corrections are also applied when converting galaxy apparent magnitudes to rest-frame absolute magnitudes, following the prescription in \cite{Kochanek_2001}.

\begin{figure}
	\centering
        \subfigure[$m_{th}$ map for $n_{\rm map}$ = 32. The sky is divided into 12,288 pixels, each covering an area of 3.36 \sqdeg.]{\includegraphics[width=0.45\textwidth]{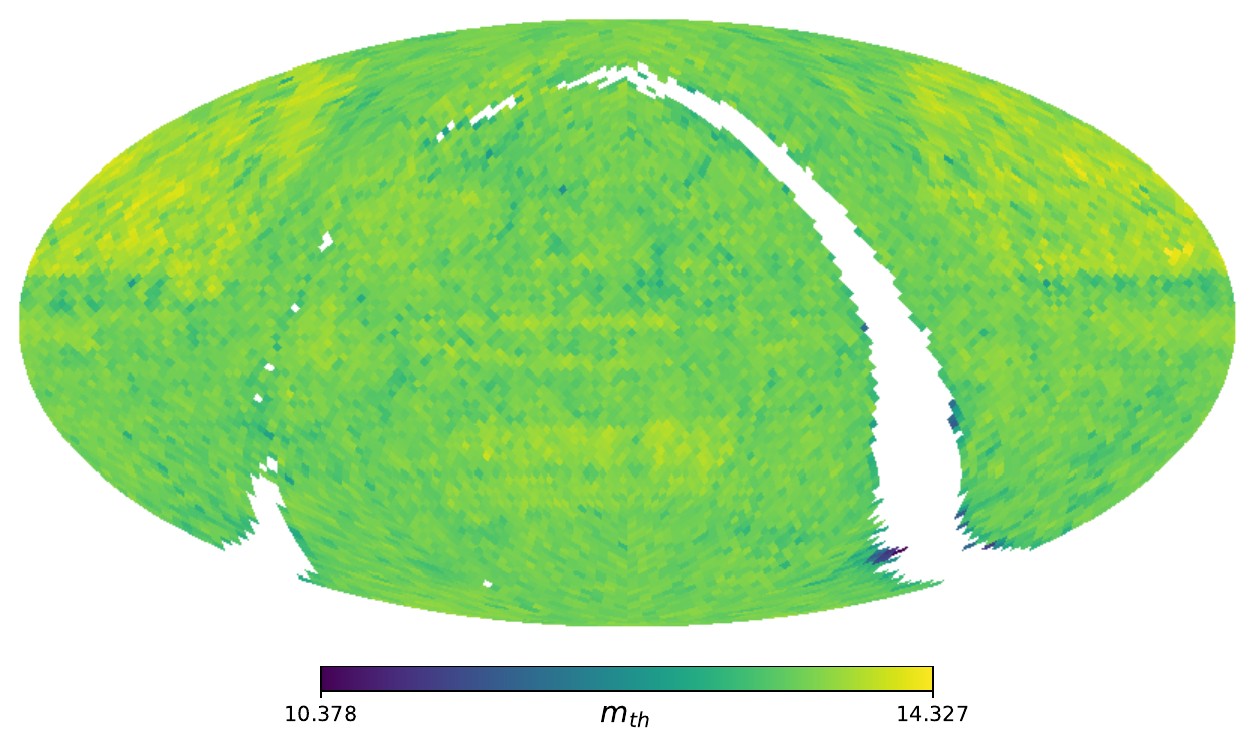}}
        \hspace{0.8cm}
        \subfigure[$m_{th}$ map for $n_{\rm map}$ = 64. The sky is divided into 49,152 pixels, each covering an area of 0.84 \sqdeg.]      {\includegraphics[width=0.45\textwidth]{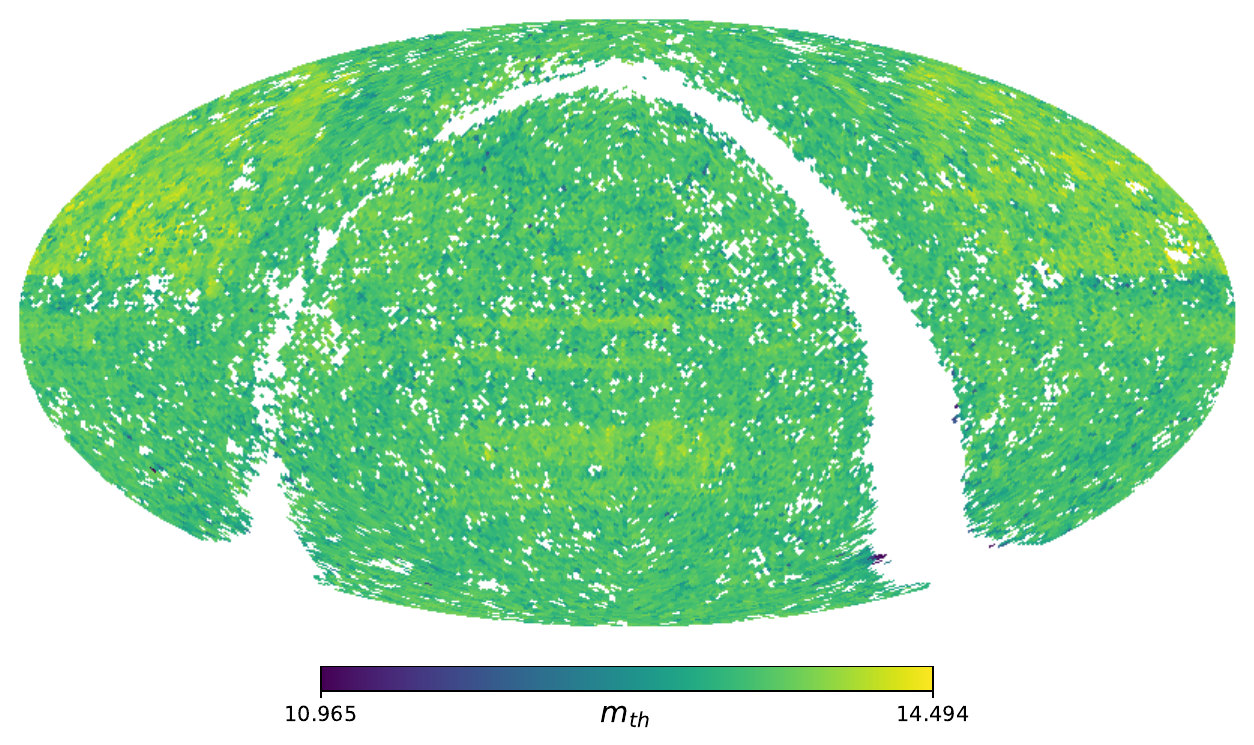}}
	\caption{$m_\text{th}$ maps for different $n_{\rm map}$ values. The minimum number of galaxies a pixel must contain in order to not be considered empty is 10. Empty pixels are shown in white.}
\label{fig: resolution choices mth maps}
\end{figure}

\begin{figure}
	\centering
        \subfigure[$m_{th}$ map for $n_{\rm map}$ = 32.]{\includegraphics[width=0.45\textwidth]{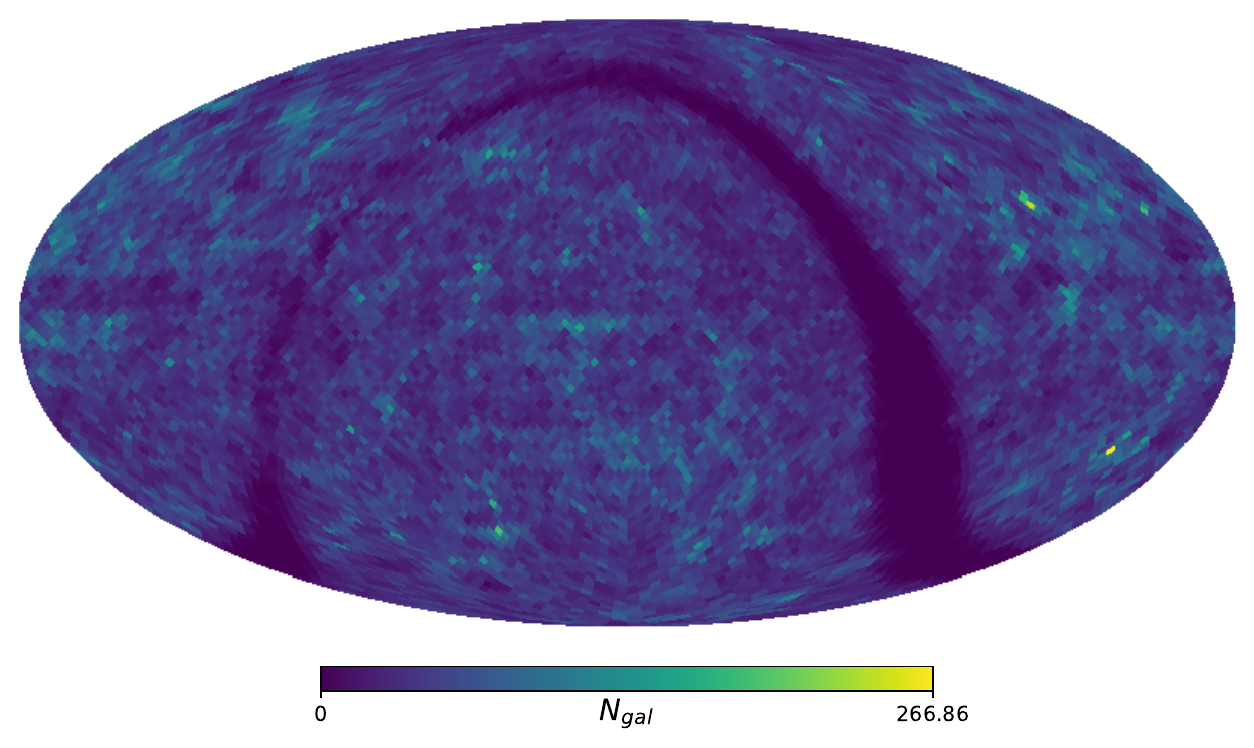}}
        \hspace{0.8cm}
        \subfigure[$m_{th}$ map for $n_{\rm map}$ = 64.]      {\includegraphics[width=0.45\textwidth]{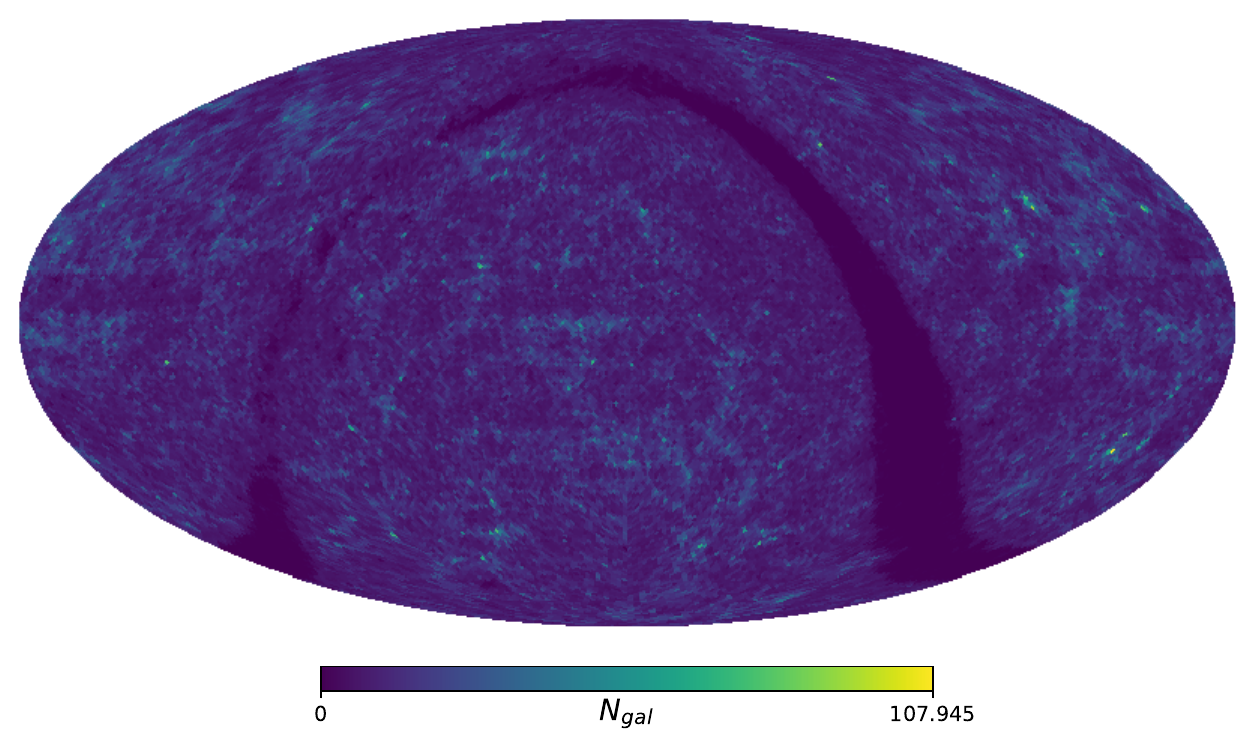}}
	\caption{Effective $N_{\text{gal}}$ maps for different $n_{\rm map}$ values.}
\label{fig: resolution choices ngal_eff maps}
\end{figure}

With the galaxy catalogue selected, and the observation band chosen, the \ac{LOS} redshift prior can now be pre-computed. The first stage of this is to pre-compute maps describing the \ac{mth} of the catalogue across the sky, as well as the corresponding numbers of galaxies. In order to assess which coarse resolution satisfies the conditions for $n_\text{map}$ listed in Sec. \ref{sec: The resolution of the LOS prior}, we have computed maps for two different $n_{\rm map}$ values: 32 and 64. These maps are shown in Figs. \ref{fig: resolution choices mth maps} and \ref{fig: resolution choices ngal_eff maps}. For the \ac{mth} maps, the minimum number of galaxies a pixel must contain in order to not be considered empty is 10. For $n_{\rm map}$=32 this results in a map for which the only empty pixels are those for which the Milky Way band obscures the sky. For $n_{\rm map}$=64, due to the limited number of galaxies with available $K$-band luminosities and the small pixel sizes, many of the pixels evaluate as empty. For this reason, $n_\text{map}=32$ is chosen to be the fiducial value (though the impact of changing to $n_\text{map}=64$ is investigated in Sec. \ref{sec:old_gwcosmo_comp}), and we do not investigate coarser resolutions, as they will not be able to accurately capture the varying galaxy catalog completeness across the sky.

With the \ac{mth} and $N_{\text{gal}}$ maps precomputed, the full \ac{LOS} redshift prior can now be computed at a higher resolution. The \ac{LOS} redshift prior (the part within square brackets of Eq. \ref{Eq:pz_withgals_sep}) is computed for every pixel assuming $n_{\rm high}$ = 32, 64 and 128. The impact of changing the choice of resolution is investigated in section \ref{sec:old_gwcosmo_comp}, and $n_\text{high}=128$ is chosen to be the fiducial value for the analyses moving forwards, as it allows the maximum information from galaxy \acp{ra} and \acp{dec} to be utilised. Higher resolutions are not investigated in this paper due to 1) the length of time it takes to compute the \ac{LOS} redshift prior (though this is parallelisable) and 2) the resulting file sizes: a \ac{LOS} prior with $n_\text{high}=128$ is about 7GB, and each subsequent increase in resolution divides each pixel into 4, and so quadruples the resulting file size. Usefully, however, this file size is completely independent of the number of galaxies which were originally in the galaxy catalogue.\footnote{Additionally, even though the file size is large, because only the pixels which cover the \eg 99.9\% sky area of the \ac{GW} event are required, the full \gwcosmo analysis does not require holding all 7GB in memory -- the data can be swiftly reduced to a manageable quantity.} The second criterion which determines the overall file size, other than pixel resolution, is the redshift resolution.

\begin{figure}
    \centering
    \includegraphics[width=0.9\textwidth]{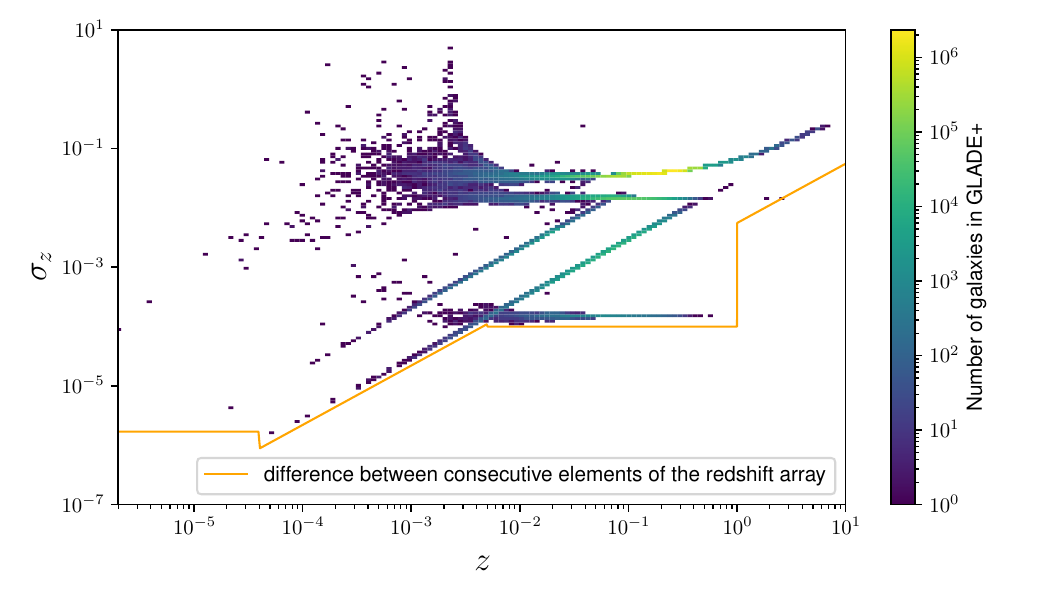}
    \caption{Scatter plot of galaxy redshift absolute uncertainty vs redshift for $K$-band galaxies in the GLADE+ catalogue, where the colour shows the number of galaxies. The spacing of the corresponding redshift array against redshift is shown in orange, demonstrating the changes in spacing required to capture all the GLADE+ galaxies with the minimum number of array elements.}
    \label{fig:zarray}
\end{figure}

Every pixel in the \ac{LOS} redshift prior is computed for the same redshift array, in order to facilitate fast operations summing pixels at later stages. For this reason, the redshift array resolution must be high enough to capture the redshift profile of every galaxy in the catalogue while maintaining an output file with manageable size. Looking at Fig. \ref{fig:zarray}, which shows a scatter plot of absolute redshift uncertainty against redshift for all the galaxies in GLADE+, it is clear that there are several distinct regions. These correspond to the different galaxy surveys which make up GLADE+, for which different uncertainty models were used.\footnote{Some galaxies can be seen to have $\sigma_z > z$; these galaxies make up a minority of the sample ($\sim 0.7\%$) and are an artefact of the photometric redshift models which were used.} We identified 4 redshift ranges in GLADE+ where the galaxies with the smallest $z$ uncertainties ($\sigma_z$) show differing dependence on $z$ and we construct the grid accordingly, shown in orange. The grid starts at $z=10^{-6}$ and ends at $z=10$, switching between linear and logarithmic spacing in order to closely match the data. By setting a requirement of having at least 1 data point in the range $[z,z+\sigma_z]$ for each galaxy, the resolution is chosen in such a way that is guaranteed to capture all of the galaxy structure in the catalogue. This does mean that in the future, as redshift uncertainties for galaxy surveys decrease, the resolution will need to increase in order to ensure this information is captured.

The choice to use the GLADE+ catalogue, rather than a deeper galaxy survey, is motivated in part so that the results which follow are directly comparable to \cite{GWTC-3:cosmology}, and partially because, by the method's construction, the same \ac{LOS} prior should be used for every \ac{GW} event. Using GLADE+ prioritises the catalogue's broad sky coverage, at the expense of greater depth in other parts of the sky. It is known that the \ac{DES} \cite{2005astro.ph.10346T} covers the footprints of GW190814 and GW170814 (two well-localised binary mergers) to a much higher level of completion, and will thus produce a more informative result for those events than the GLADE+ catalogue \cite{Soares-Santos:2019irc,GW190814:DES}. However, the inclusion of these deeper surveys is left to future work.


\section{Results \label{sec:results}}

We choose to use the same fiducial population models as \cite{GWTC-3:cosmology}. For mass models, the \textsc{Power Law + Peak} mass distribution is used for \ac{BBH}. The distribution of the primary mass, $m_1$, is given by
\begin{equation}\label{eq:powerlaw+peak}
p(m_1|M_\text{min}^\text{BH}, M_\text{max}^\text{BH}, \alpha, \lambda_g, \mu_g, \sigma_g) = [(1 - \lambda_g) \mathcal{P}(m_1|M_\text{min}^\text{BH}, M_\text{max}^\text{BH}, -\alpha) + \lambda_g \mathcal{G}(m_1|\mu_g, \sigma_g)],
\end{equation}
while the secondary mass, $m_2$ is described by a truncated power-law with slope $\beta$ between $M_\text{min}^\text{BH}$ and $m_1$.
The \ac{NSBH} events also use the \textsc{Power Law + Peak} mass distribution for the primary mass, but with the secondary mass using a uniform prior between $M_{\text{min}}^\text{NS}$ and $\min(M_{\text{max}}^\text{NS},m_1)$. The \acp{BNS} use a uniform prior between $M_{\text{min}}^\text{NS}$ and $M_{\text{max}}^\text{NS}$. Table \ref{tab:params_table} describes the parameters used in the mass distributions and gives the prior ranges which are used for the analyses in Secs. \ref{sec:icaro_comp} and \ref{sec:pop+catalog_results}.

\begin{table}
\small
\centering
POWER LAW + PEAK \\
\vspace{2pt}
\begin{tabularx}{\textwidth}{ C{0.4} L{2.0} L{0.6} } 
 \hline
 \bf{Parameter}  & \bf{Description} & \bf{Prior}    \\ 
 \hline\hline
 $M_{\text{min}}^\text{BH}$ & Minimum mass of the PL component of the black hole mass distribution. & $\mathcal{U}(2.0  $M$_{\odot},10.0 $M$_{\odot})$  \\ 
 $M_{\text{max}}^\text{BH}$ & Maximum mass of the PL component of the black hole mass distribution. & $\mathcal{U}(50.0  $M$_{\odot},200.0 $M$_{\odot})$  \\
 $\alpha$ & Spectral index for the PL of the primary mass distribution. & $\mathcal{U}(1.5,12.0)$ \\
 $\mu_g$ & Mean of the Gaussian component in the primary mass distribution. & $\mathcal{U}(20.0  $M$_{\odot},50.0 $M$_{\odot})$ \\
 $\sigma_g$ & Width of the Gaussian component in the primary mass distribution. & $\mathcal{U}(0.4  $M$_{\odot},10.0 $M$_{\odot})$ \\
 $\lambda_g$ & Fraction of the model in the Gaussian component. & $\mathcal{U}(0.0,1.0)$ \\
 $\delta_m$ & Range of mass tapering on the lower end of the mass distribution. & $\mathcal{U}(0.0  $M$_{\odot},10.0 $M$_{\odot})$ \\
 $\beta$ & Spectral index for the PL of the secondary mass distribution. & $\mathcal{U}(-4.0,12.0)$ \\
 $M_{\text{min}}^\text{NS}$ & Minimum mass of the uniform neutron star mass distribution. & $\mathcal{U}(1.0$M$_{\odot}, 1.5$M$_{\odot})$  \\ 
 $M_{\text{max}}^\text{NS}$ & Maximum mass of the uniform neutron star mass distribution. & $\mathcal{U}(2.0$M$_{\odot}, 5.0$M$_{\odot})$ \\
 \hline
 \end{tabularx}
\caption{Summary of mass distribution parameters with the corresponding prior ranges.}
\label{tab:params_table}
\end{table}

The \ac{CBC} merger rate evolution model is assumed to follow a Madau-Dickinson-like distribution~\citep{Madau_2014} which is given by:
\begin{equation}\label{eq:madau-diskinson}
    R(z) = R_0(1+z)^\gamma\,\frac{1+ (1+z_p)^{-(\gamma+\kappa)}}{1+\left(\frac{1+z}{1+z_p}\right)^{\gamma+\kappa}}.
\end{equation}
Definitions of the parameters appearing in Eq. \ref{eq:madau-diskinson} are given in Tab. \ref{tab:merger_params_table}, as well as the prior ranges which are used for the analyses in Secs. \ref{sec:icaro_comp} and \ref{sec:pop+catalog_results}.

\begin{table}
\small
\centering
MERGER RATE SHAPE PARAMETERS
\vspace{2pt}
\begin{tabularx}{\textwidth}{ C{0.4} L{2.0} L{0.6} } 
 \hline
 \bf{Parameter}  & \bf{Description} & \bf{Prior}    \\ 
 \hline\hline
 $R_0$ & Local merger rate. & $1/R_0$ (implicit) \\
 $\gamma$ & Power-law index describing the merger rate at low redshift. &  $\mathcal{U}(0,12.0)$\\
 $\kappa$ & Power-law index describing the merger rate at high redshift. & $\mathcal{U}(0,6.0)$ \\
 $z_p$ & The redshift where the slope of the merger rate changes.  &$\mathcal{U}(0,4.0)$\\
 \hline
\end{tabularx}
\caption{Summary of merger rate shape parameters with the corresponding priors.}
\label{tab:merger_params_table}
\end{table}

Using the selection of events described in \ref{sec:GWdata}, three separate analyses are carried out: a pure population analysis using 42 \acp{BBH} (see Sec. \ref{sec:icaro_comp}), a galaxy catalogue analysis with fixed population assumptions using all the \acp{BBH}, \acp{NSBH} and \acp{BNS} (see Sec. \ref{sec:old_gwcosmo_comp}) and, finally, a population + galaxy catalogue analysis which uses all \ac{GW} events and galaxy catalogue information, and jointly estimates cosmological and population parameters (Sec. \ref{sec:pop+catalog_results}). For the multi-parameter estimation analyses, \gwcosmo utilises nested sampling with the sampler \dynesty \cite{2020MNRAS.493.3132S},\footnote{\url{https://zenodo.org/record/7995596}} while for the \ac{H0}-only analysis a gridded method is used.

\subsection{The population analysis of GWTC-3 BBHs \label{sec:icaro_comp}}

\begin{figure}
    \centering
    \includegraphics[width=0.95\textwidth]{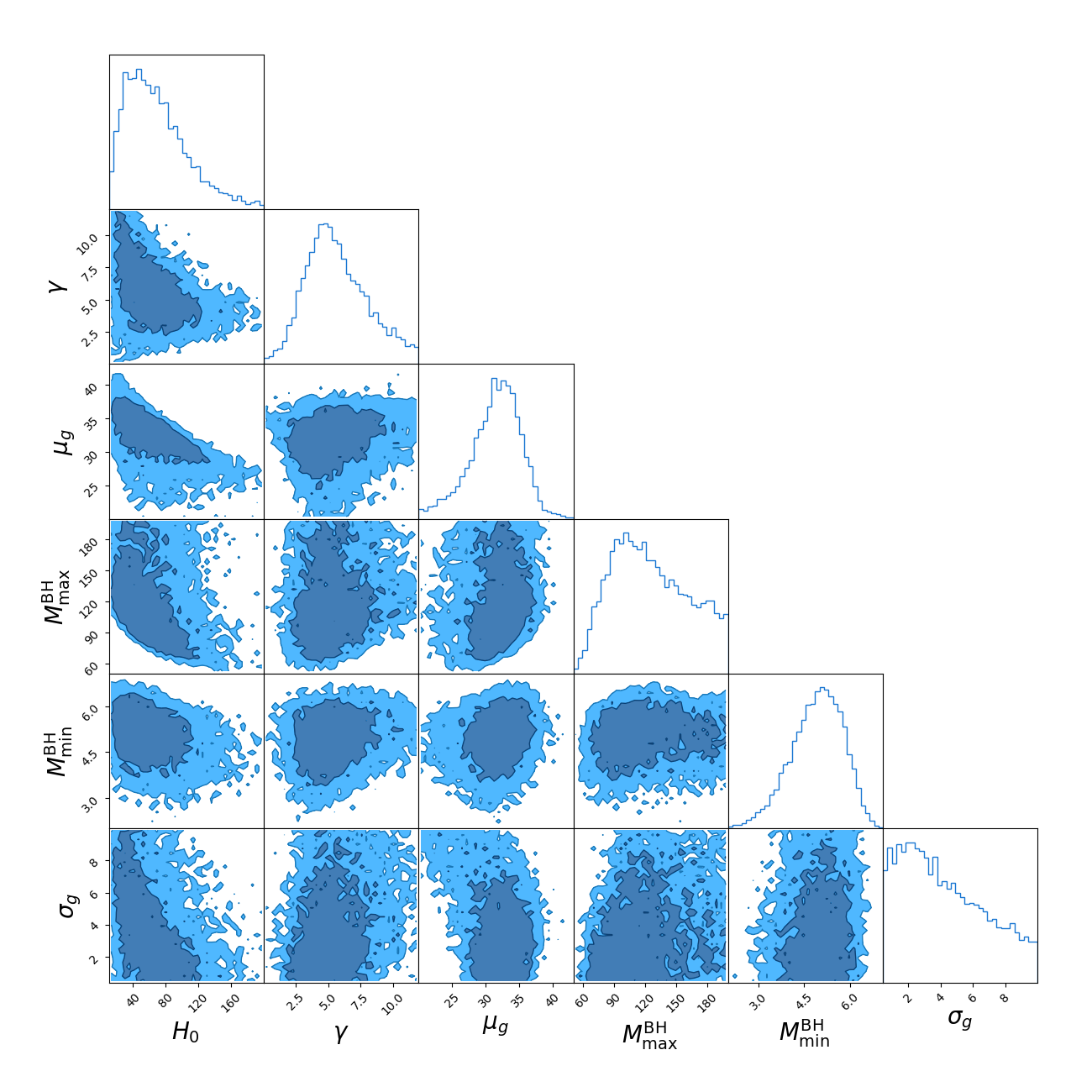}
\caption{Posteriors on parameters $H_0$, $\gamma$, $\mu_g$, $M_\text{max}^\text{BH}$, $M_\text{min}^\text{BH}$ and $\sigma_g$ using the 42 BBH events of the GWTC-3 catalog, obtained with a pure population analysis with \gwcosmo.}\label{fig:gwcosmo_pop}
\end{figure}

In this section the 42 \acp{BBH} are reanalysed using \gwcosmo, allowing cosmological and population parameters to vary. Fig.~\ref{fig:gwcosmo_pop} shows the result of the analysis for \ac{H0} and the parameters which correlate most strongly with it, namely $\mu_g, M_\text{min}^\text{BH}, M_\text{max}^\text{BH}$ and $\sigma_g$ from the mass model (see Table~\ref{tab:params_table}) and $\gamma$ from the merger rate model (see Table~\ref{tab:merger_params_table}).  The full corner plot showing all parameters can be found in appendix \ref{app:full corner}. Marginalising over population parameters results in a value of $H_0=$\BBHpopempty (maximum a posteriori probability (MAP) and 68\% highest density interval (HDI)). All results from this point on will be quoted using the same, unless specified otherwise. The result is fully compatible with the \icarogw result from \cite{GWTC-3:cosmology}, with minor deviations due to the versions of \ac{GW} parameter estimation which were used in each analysis (\cite{GWTC-3:cosmology} used parameter estimation from the GWTC-2 release, whereas we make use of the updated GWTC-2.1 release), and the fact that \gwcosmo implicitly marginalises over $R_0$ with a $1/R_0$ prior, while in \cite{GWTC-3:cosmology} \icarogw used a uniform prior on $R_0$.

\subsection{The galaxy catalogue analysis of GWTC-3 with fixed population assumptions \label{sec:old_gwcosmo_comp}}

\begin{figure}
    \centering
    \includegraphics[width=0.92\textwidth]{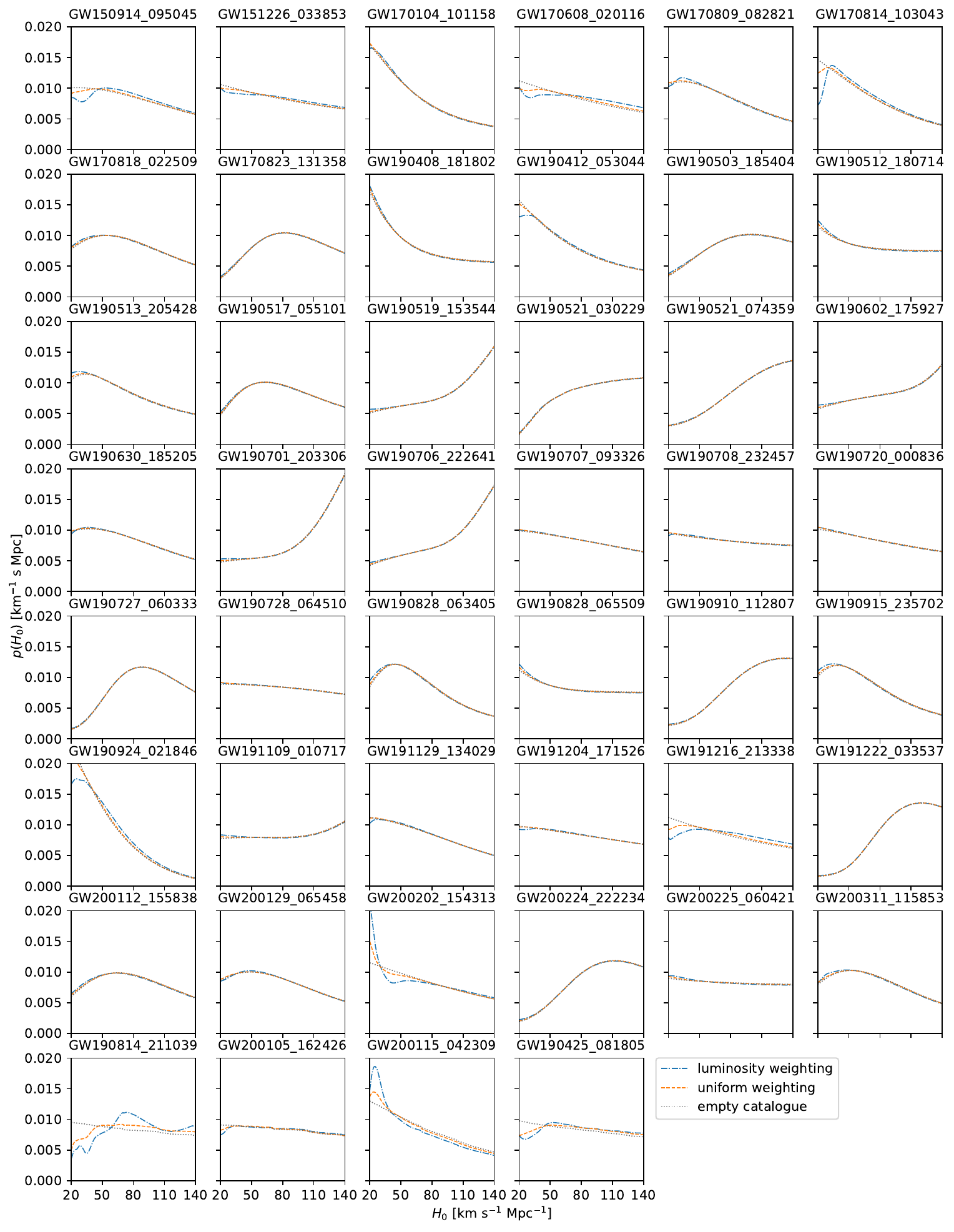}
\caption{$H_0$ posteriors (uniform prior) for the 46 dark sirens from GWTC-3 with \ac{SNR} $>11$. The dotted line shows the contribution coming purely from the \ac{GW} population assumptions and no galaxy catalogue information (the ``empty catalogue'' scenario). The dashed orange line includes galaxy catalogue information with uniform weighting of host galaxies, while the blue dot-dashed line includes galaxy catalogue information and applies weighting proportional to the galaxy's luminosity.}
\label{fig: H0 posteriors all events}
\end{figure}

In order to best assess the individual contributions of \ac{GW} events to the \ac{H0} measurement, as well as determine to what extent a result is driven by population information versus galaxy catalogue information, it is useful to run an analysis with fixed population assumptions. This analysis is carried out for the 46 dark siren events from GWTC-3 with \ac{SNR} $> 11$. For the \textsc{Power Law + Peak} mass model we use $\alpha = 3.78$, $\beta = 0.81$, $M_{\text{max}}^\text{BH} = 112.5 M_\odot$, $M_{\text{min}}^\text{BH} = 4.98 M_\odot$ , $\delta_m = 4.8 M_\odot$ $\mu_g = 32.27 M_\odot$ , $\sigma_g = 3.88 M_\odot$ and $\lambda_g = 0.03$. For neutron stars we fix $M_{\text{min}}^\text{NS} = 1M_\odot $and $M_{\text{max}}^\text{NS} = 3M_\odot$. For the evolving merger rate model we use $\gamma = 4.59$, $k = 2.86$ and $z_p = 2.47$. These values match those used for the \gwcosmo analysis in \cite{GWTC-3:cosmology}, using the previous version of \gwcosmo, making the results directly comparable.

The posteriors for individual events (assuming uniform priors on \ac{H0}) can be seen in Fig. \ref{fig: H0 posteriors all events}, for three different analysis assumptions: luminosity weighting of host galaxies in the $K$-band, uniform weighting of host galaxies, and no galaxy catalogue (also known as the ``empty catalogue'' analysis). The empty catalogue analysis demonstrates the information which is driven by the chosen \ac{GW} population, and as such it is the difference between this line and the others which should be used to assess how informative the galaxy catalogue method is for any given event. For the majority of events, lying at distances higher than that to which the GLADE+ catalogue is complete in the $K$-band, the results of all 3 analyses are the same. Some nearer-by events show deviations at the low-\ac{H0} end, which corresponds to low redshift (for a fixed luminosity distance) where the galaxy catalogue is more complete. In general, the luminosity weighted results are the most informative, as the majority of the \ac{GW} host probability gets condensed into a smaller number of galaxies, while in the unweighted case the presence of many low-luminosity galaxies means that the final \ac{LOS} redshift prior will appear much closer to uniform in comoving volume when averaged over the volume of an event.

\begin{figure}
	\centering    
        {\includegraphics[width=0.65\textwidth]{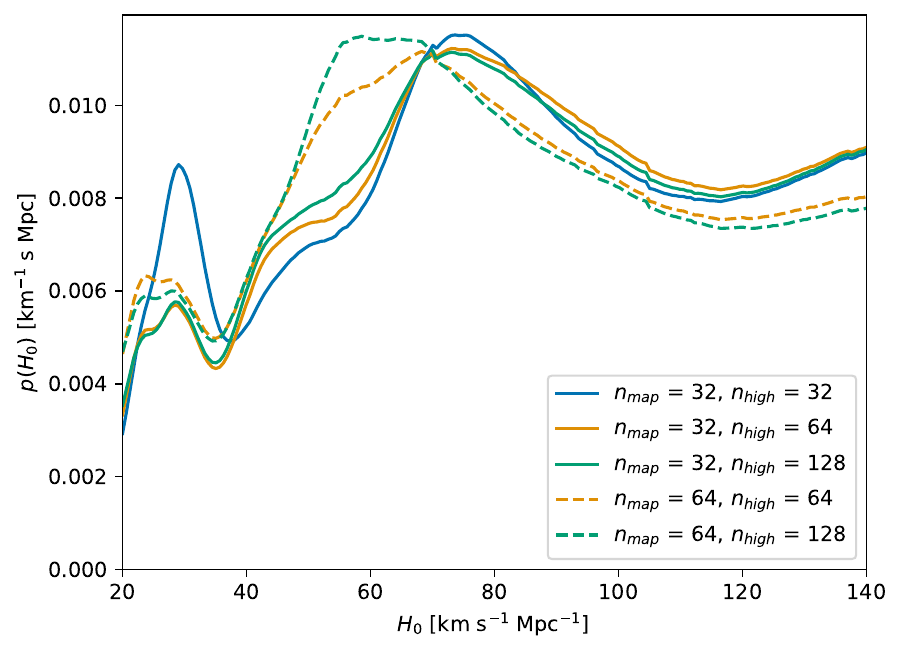}}
	\caption{$H_0$ posteriors for GW190814 using different values for $n_{\rm high}$ and $n_{\rm map}$.}
\label{fig: GW190814 H0 posteriors}
\end{figure}

\begin{figure}
    \centering
    \includegraphics[width=0.65\textwidth]{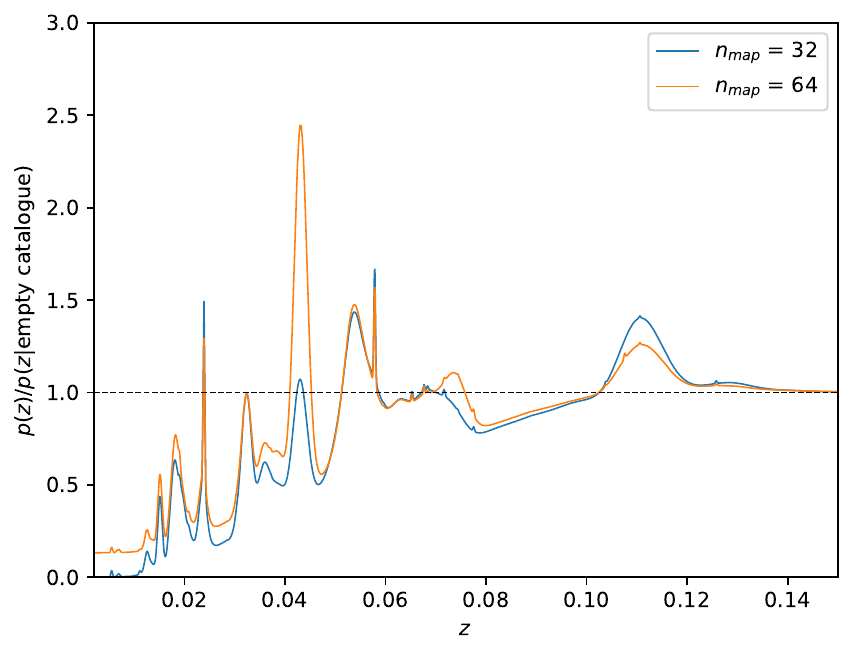}
\caption{Ratio of the galaxy catalogue and empty catalogue LOS redshift prior, weighted by the sky area of GW190814 with $n_{\rm high} = 128$. The blue curve corresponds to the $n_{\rm map}$=32 case and the orange curve corresponds to $n_{\rm map}$=64. The horizontal dashed black line at 1.0 corresponds to the scenario when matter is distributed truly uniformly in comoving volume, so oscillations above and below this correspond to over- and under-densities in matter. At high redshift, where the catalogue contains no information in the $K$-band, the curves tend to 1.0 as the incompleteness correction dominates.}
\label{fig: LOS zprior examples}
\end{figure}

The event of most interest, from a galaxy catalogue perspective, is GW190814.  Due to its nearby distance (around 240 Mpc) and small sky localisation (18.5 \sqdeg for the 90\% HDI) \cite{GW1908142020} this event is an ideal dark siren. In order to assess the impact of analysis assumptions relating to the choice of pixel resolution, GW190814 is reanalysed under a variety of different assumptions (see Fig. \ref{fig: GW190814 H0 posteriors}). Two different resolutions of \ac{mth} and $N_\text{gal}$ maps are used: $n_{\rm map}$=32 and 64 (see Figs. \ref{fig: resolution choices mth maps} and \ref{fig: resolution choices ngal_eff maps}), and for each of these, different resolutions up to $n_{\rm high}$=128 are used for the final \ac{LOS} prior. Interestingly, the choice of map resolution seems to have a larger impact on the shape of the posterior than that of $n_\text{high}$. Looking at Fig. \ref{fig: LOS zprior examples}, it can be seen that while the same galaxies are present in the \ac{LOS} redshift prior for the $n_{\rm map}$=32 and 64 cases, the relative contributions of these galaxies can differ significantly. The driving reason for this is the small number statistics discussed in detail in Sec. \ref{sec: The resolution of the LOS prior}: in the $n_{\rm map}$=64 scenario, a lot of pixels within GW190814's sky area only contain about a dozen galaxies, meaning that the relative contributions from adjacent pixels can be significantly impacted. The larger pixel sizes in the $n_{\rm map}$=32 scenario reduce this impact, and the results are more robust because of it.

The value of $n_\text{high}$ also impacts the final shape of the posterior. This is most noticeable for the $n_\text{high}=32$ result, compared with the $n_\text{high}=64$ and 128 results using $n_{\rm map}$=32 in all cases. A lower redshift galaxy which contributes a large peak for $n_\text{high}=32$ is subsequently suppressed at higher resolutions, because the pixel it lies within contains significant variation in \ac{GW} support: a detail which is only captured at higher resolutions. The $n_\text{high}=64$ and 128 results are very similar, showing a reasonable level of convergence has been reached. As GW190814 is the most well-localised dark siren in GWTC-3, we conclude that this resolution is sufficient to capture all the necessary \ac{GW} information for this analysis. It is worth remembering, however, that the choice of resolution which is ``good enough'' will depend on the localisation area of the \ac{GW} events it is used with -- a future \ac{GW} event with a smaller localisation region that GW190814 will need a higher resolution \ac{LOS} prior to do it justice.

\subsection{Combined cosmological and population inference using GWTC-3\label{sec:pop+catalog_results}}

With the robustness of this version of \gwcosmo established, the final step is to do a full population + galaxy catalogue analysis of GWTC-3. The \acp{BBH} are analysed as in Sec. \ref{sec:icaro_comp}, but this time utilising a \ac{LOS} redshift prior constructed from the GLADE+ $K$ band, using a resolutions of $n_\text{map}=32$ and $n_\text{high}=128$. The two \acp{NSBH} and GW190814 are also analysed using the full population + galaxy catalogue method. The maximum and minimum mass of the \ac{NS} component is allowed to vary in this analysis, as GW190814's secondary mass is very heavy for a \ac{NS} and we want to avoid a cut off in the mass distribution which could incorporate additional cosmological information into the analysis (see appendix~\ref{app:full corner} for the full results of this analysis). When considering these 3 events alone, the majority of population parameters are poorly constrained (as expected, see also the NSBH analysis in \cite{AChen2023}). Because GW190425 is the only \ac{BNS} without an \ac{EM} counterpart we do not carry out a population analysis for it. Given the lack of features in the assumed BNS mass distribution (assuming the event is not so heavy or light as to touch the BNS prior bounds), we simply fix the population parameters to fiducial values and carry out a 1-dimensional analysis over \ac{H0}.

\begin{figure}
    \centering
    \includegraphics[width=0.95\textwidth]{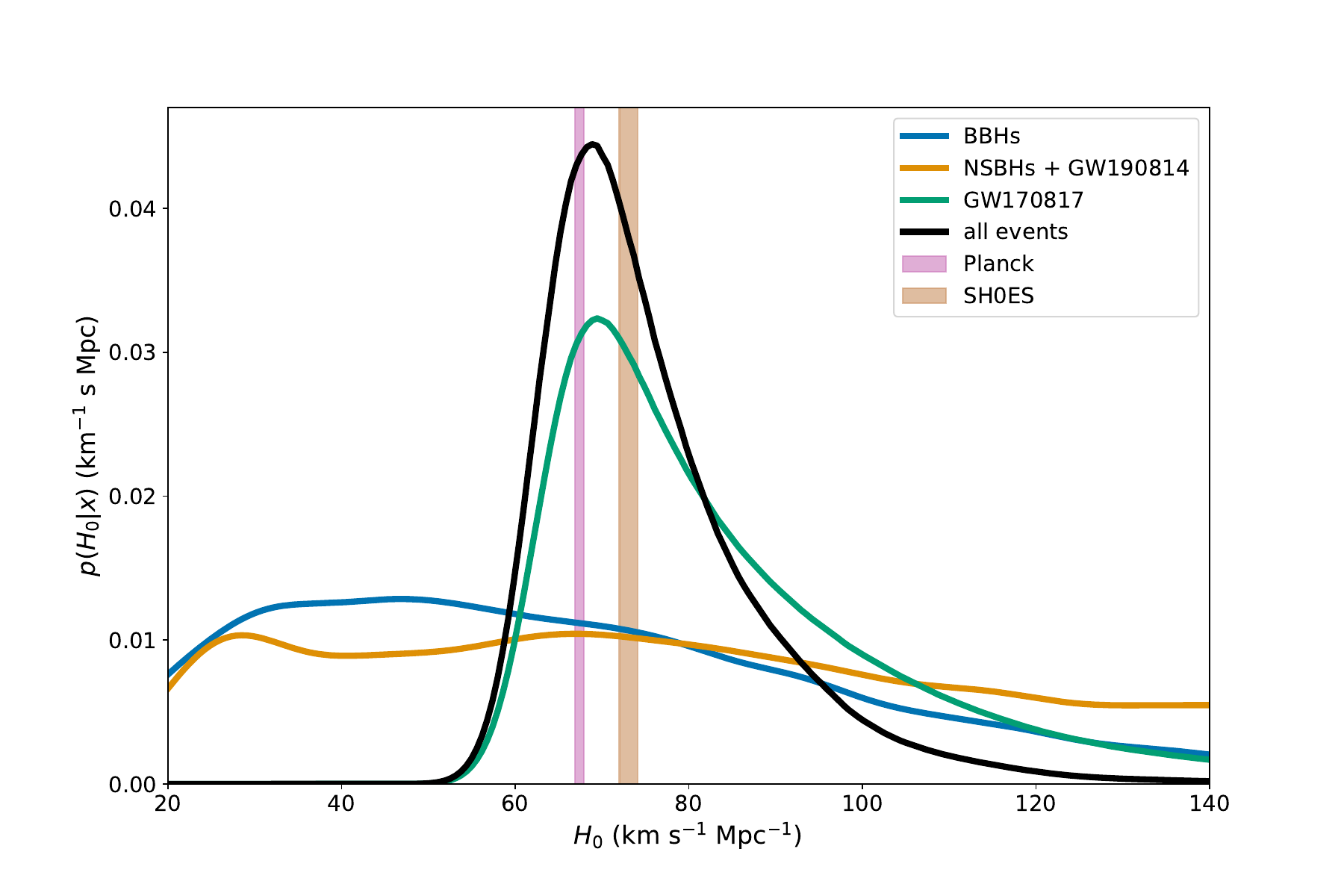}
\caption{Posterior on $H_0$ produced using 42 BBHs, 2 NSBHs, GW190814, and GW190425 using the GLADE+ galaxy catalogue and GW170817 with its \ac{EM} counterpart (black line). The separate contributions from the BBHs (blue) the NSBHs and GW190814 (orange) and GW170817 (green) are also shown (normalised). The $1\sigma$ Planck \cite{Aghanim:2018eyx} and SH0ES \cite{Riess_2022} results are shown as shaded pink and brown bands respectively.}\label{fig:combinedH0}
\end{figure}

The result which combines the inference from all 3 GW populations is presented in Fig.~\ref{fig:combinedH0}. The final constraint on \ac{H0} is \combinedpoppluscatalogue which, as anticipated, is less informative than the previous result produced by \gwcosmo in \cite{GWTC-3:cosmology} of $H_0=$ \oldgwcosmo, due to the marginalisation over the uncertainty of the population hyper-parameters. While less informative, this result is far more robust, and no longer susceptible to bias due to fixed assumptions about the \ac{GW} population.


\section{Conclusions\label{sec:conclusions}}

This paper presents a novel method for cosmological analyses using \ac{GW} data and galaxy catalogues, incorporating information from the \ac{GW} mass distribution in a way which allows the joint estimation of cosmological and \ac{GW} population parameters, a crucial improvement while the true mass distribution of compact mergers remains uncertain. A key part of this new method is the construction of a \ac{LOS} redshift prior which incorporates both galaxy catalogue data and an incompleteness correction, the pre-computation of which produces a speed-up in the analysis of roughly 3 orders of magnitude over the previous version of \gwcosmo. Comparing the population-only cosmological results using 42 \acp{BBH} from GWTC-3 to the \icarogw results in \cite{GWTC-3:cosmology} shows excellent agreement between the methods. Similarly, the fixed population result using the GLADE+ galaxy catalogue agrees very well with the previous version of \gwcosmo in \cite{GWTC-3:cosmology}. The ability to jointly estimate, or marginalise over, the population of compact mergers means that this analysis is largely robust to the current uncertainties in the population. The LVK's fourth observing run (O4) is currently underway and is expected to increase the number of \ac{GW} detections to several hundreds. This improved version of \gwcosmo will allow these new events to be used in an informative cosmological analysis which incorporates data from current galaxy surveys, and also utilises information from the \ac{GW} population itself. Nearby and well localised sources will make informative contributions due to the known galaxy structure within their localisation volumes. The numerous \ac{GW} events at high distances, where catalogue support is low will still contribute cosmologically through their redshifted masses. By joining these two sources of redshift information the best of both is utilised, enabling reliable and informative future measurements of \ac{H0} using dark sirens.

However, there are still areas in which further improvements must be made. Here, for example, in the final result we treat population hyper-parameters as uncorrelated between the different source populations. Ideally, parameters which are common between the populations (\eg the hyper-parameters describing the black hole as the primary mass in both the BBH and NSBH cases) would be estimated jointly. However for certain parameters, such as those parameterising the merger rate evolution model, one would not expect to be the same between the different \ac{GW} populations. There, and in general perhaps, the solution could be to infer hyper-parameters for the entire \ac{CBC} population, rather than breaking it into sub categories. Given the current difficulties in firmly placing current \ac{GW} detections into these categories (GW190814 being an excellent example, not to mention the current uncertainty surrounding the existence -- or not -- of a mass gap between the NS and BH population \cite{PhysRevX.13.011048}) this could simplify the process noticeably.

While the \gwcosmo method is now able to jointly constrain cosmological and population hyper-parameters, it shouldn't be forgotten that the choice of mass model can have an impact on the inference, and ensuring that a model is used which adequately captures the structure in the compact binary mass distribution will be an important aspect of this analysis moving forwards. Similarly, while current evidence does not confirm whether the mass distribution of compact binaries evolves with redshift or remains constant, this would be an important feature to include should we reach the point where the data shows a strong preference one way or the other.

From the galaxy catalogue side of things, the current \gwcosmo methodology is much improved to handle future data releases, as the speed of the analysis is not tied to the number of galaxies contained within the catalogue. It can also be adapted with much more ease in future to non-gaussian redshift uncertainties, which may be of high importance in the regime where the galaxy catalogue information becomes a more dominant contributor to the measurement of \ac{H0}. With current galaxy surveys now publicly available which are complete to higher redshifts than the GLADE+ catalogue (such as the DESI Legacy Imaging Surveys \cite{2019AJ....157..168D} and DES \cite{2005astro.ph.10346T}), and upcoming surveys promising ever greater depths and redshift precision (such as the DESI survey \cite{2016arXiv161100036D}, LSST \cite{2019ApJ...873..111I} and Euclid \cite{2022A&A...662A.112E}) the era of galaxy catalogue driven dark siren analyses is just about to begin. 

The main results of this paper, $H_0 =$ \combinedpoppluscatalogue, provides a robust measurement of the Hubble constant which combines information from GW events in combination with \ac{EM} counterparts, galaxy catalogues, and redshifted mass information, paving the way forward for future GW cosmology analyses which can utilise information from an array of different sources. With the number of \ac{GW} events set to vastly increase over the coming years, and the galaxy survey information becoming available, this is a truly exciting time for the field of gravitational-wave cosmology.

\begin{acknowledgments}
The authors would like to thank Jonathan Gair and Sylvain Marsat for their contributions to reviewing the \gwcosmo code, as well as Danny Laghi and Suvodip Mukherjee for their useful comments on the paper draft.

This material is based upon work supported by NSF's LIGO Laboratory which is a major facility fully funded by the National Science Foundation. The authors are grateful for computational resources provided by the LIGO Laboratory and supported by National Science Foundation Grants PHY-0757058 and PHY-0823459. The research of RG was suported by ERC starting grant SHADE 949572 and STFC grant ST/V005634/1. FB, CT and AG are supported by the Ghent University Special Research Funds (BOF) project BOF/STA/202009/040 and the Fonds Wetenschappelijk Onderzoek (FWO) iBOF project BOF20/IBF/124. A.C. is supported by a PhD grant from the Chinese Scholarship Council (grant no.202008060014). T.B. is supported by ERC Starting Grant \textit{SHADE} (grant no.~StG 949572) and a Royal Society University Research Fellowship (grant no.~URF$\backslash$R1$\backslash$180009). A.E.R. was supported by the UDEA Dedicacion exclusiva and Sostenibilidad programs and the CODI projects 2019-28270 and 2021-44670 of UDEA.
\end{acknowledgments}

\appendix

\section{Full population results \label{app:full corner}}
Here we present the full parameter constraints resulting from the analyses described in section \ref{sec:results}.

\begin{figure}
    \centering
    \includegraphics[width=1.\textwidth]{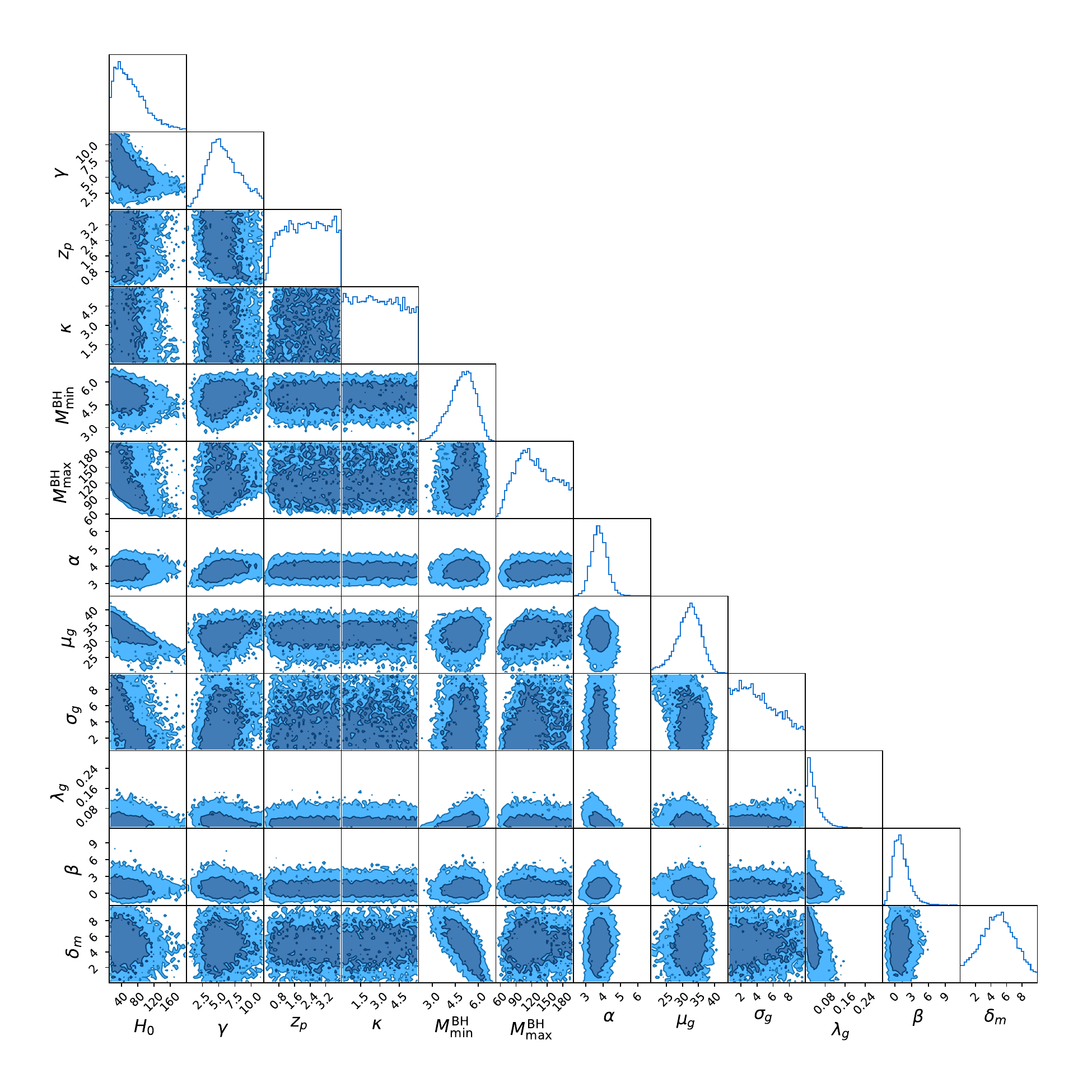}
\caption{Posteriors on 12 cosmological and population parameters using the 42 BBH events of the GWTC-3 catalog, obtained with a pure population analysis with \gwcosmo.}\label{fig:gwcosmo_pop_full}
\end{figure}

\begin{figure}
    \centering
    \includegraphics[width=1.\textwidth]{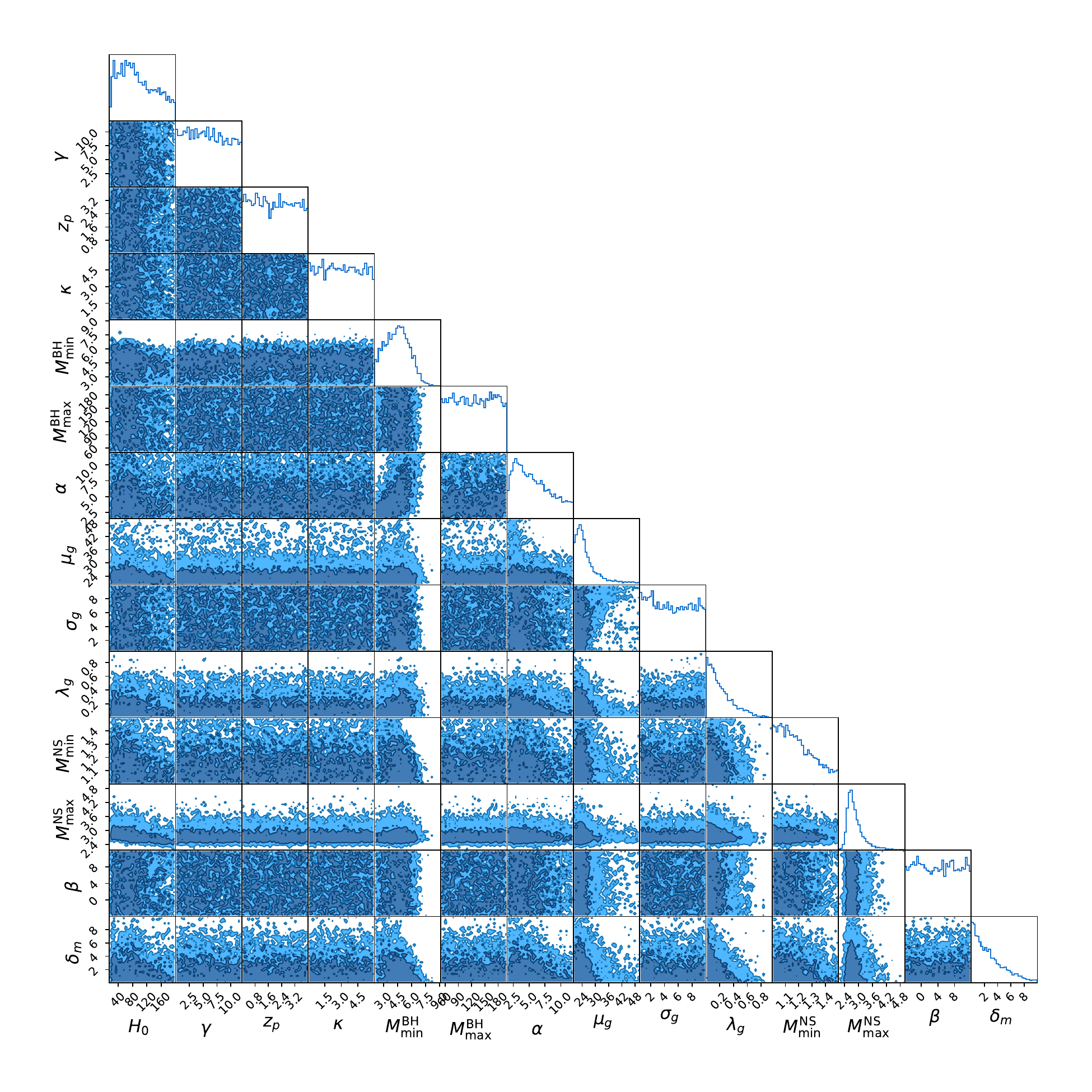}
\caption{Posteriors on 14 cosmological and population parameters using the 2 NSBH events (GW200105 and GW200115) and GW190814, including information from the GLADE+ galaxy catalogue.}\label{fig:nsbh_full_corner}
\end{figure}

\bibliographystyle{JHEP}
\bibliography{masterbib}

\end{document}